\documentclass[11pt,reprint,showkeys, amsmath,amssymb, aps, floatfix]{revtex4-2}
\DeclareUnicodeCharacter{2212}{\textendash}
\usepackage{graphicx,subfigure}
\graphicspath{{figures/}}
\usepackage[export]{adjustbox}
\usepackage{dcolumn}
\usepackage[outdir=./epsTopdf/]{epstopdf}
\usepackage{bm}
\usepackage{amsmath}
\usepackage{amssymb}
\usepackage{physics}
\usepackage{xcolor}
\usepackage{slashed,cancel}
\usepackage{multirow}
\usepackage{float}
\usepackage{capt-of}
\usepackage{comment}
\usepackage{ mathrsfs }
\usepackage{enumitem}
\usepackage[colorlinks=true,linkcolor=blue,urlcolor=blue,citecolor=blue]{hyperref}

\begin{document}
	
	\preprint{APS/123-QED}

	\title{Anomalous triple gauge couplings in $e^-e^+\to 4j$: Role of polarizations, spin correlations and interference }
	
	\author{Amir Subba}
	\email{as19rs008@iiserkol.ac.in}
	
	\author{Ritesh K. Singh}
	\email{ritesh.singh@iiserkol.ac.in}
	\affiliation{Department of Physical Sciences, Indian Institute of Science Education and Research Kolkata, Mohanpur, 741246, India\\
	}
	
	\date{\today}
	
	\begin{abstract}
		We investigate the anomalous charged triple gauge boson couplings generated via $SU(2)_l\times U(1)_Y$ gauge invariant dimension-6 operators with final state four jet events in an $e^-e^+$ Collider at $\sqrt{s}=250$~GeV. We consider all the leading order contributions including the contribution from the interference of $W^-W^+$ diagrams with other possible diagrams. The tagging of two $W$ bosons with a pair of jets is done using the jet charge, while the decay products of $W^\prime$s are tagged as \emph{up/down}-type using boosted decision trees to construct polarizations and spin correlations. Marginalized limits on five anomalous couplings are obtained by Markov Chain Monte Carlo analysis using polarizations, spin correlations, and cross~section. We found that the full-hadronic channel provides tighter limits on anomalous couplings than compared to the usually sought-after clean semi-leptonic channel, owing to the large contribution from interference. 
	\end{abstract}
	
	
	\maketitle
	
	\section{Introduction}
	The Standard Model~(SM) of particle physics is a theoretical framework that combines quantum mechanics, special relativity, and gauge symmetry to accurately describe the fundamental particles and their interactions. It is based on the gauge symmetry group $SU(3)_C\times SU(2)_L\times U(1)_Y$, which corresponds to the color, and electroweak interaction. One of the crucial aspects of the SM is the existence of the Higgs boson, which was discovered by the CMS~\cite{CMS:2012qbp} and ATLAS~\cite{ATLAS:2012yve} collaborations at the Large Hadron Collider~(LHC). The Higgs boson plays a fundamental role in the SM by providing mass to all the elementary particles through a Higgs mechanism~\cite{Higgs:1964pj,Englert:1964et}. The non-abelian gauge structure of SM allows for charged triple gauge couplings~$(W^-W^+\gamma/Z)$ in electroweak sector. The $W^-W^+\gamma/Z$ structure provides a unique platform to probe for any deviation, should any physics beyond SM~(BSM) is to exist. These coupling are experimentally probed in $W^-W^+$~\cite{ATLAS:2021jgw,CMS:2013ant,ATLAS:2012upi}, $W^\pm\gamma/Z$~\cite{CMS:2021icx,CMS:2019efc,ATLAS:2016qjc,ATLAS:2016bkj,CMS:2013ryd,ATLAS:2013way,ATLAS:2012bpb,CDF:2012mnr} di-boson processes. The search for deviation from SM prediction is fueled by many incompleteness of SM to explain stability of Higgs mass in presence of quantum corrections, structure of dark matter, neutrino masses, etc. Experiments at hadronic colliders~(such as LHC and Tevatron)~\cite{ATLAS:2021jgw,CMS:2013ant,ATLAS:2012upi,CMS:2021icx,CMS:2019efc,ATLAS:2016qjc,ATLAS:2016bkj,CMS:2013ryd,ATLAS:2013way,ATLAS:2012bpb,CDF:2012mnr} have probed the anomalous contribution to $W^-W^+\gamma/Z$ couplings in semi-leptonic, and full-leptonic final states, owing to an ideal reconstruction of final states.\\
	The large branching ratio of $W^\pm/Z$ boson to decay in a hadronic states provides a significant sensitive phase space to new physics. In this article, we study $W^-W^+$ process in four jets final states to probe anomalous behavior in $W^-W^+\gamma/Z$ couplings at polarized $e^-e^+$ collider.
	The most general $W^-W^+\gamma/Z$ couplings are usually parameterize in terms of $14$ parameters as~\cite{Hagiwara:1986vm},
	\begin{equation}
		\label{eqn:eff}
		\begin{split}
			&\mathscr{L}_{WWV} = ig_{WWV}\left[g_1^V(W^+_{\mu \nu}W^{-\mu} - W^{+\mu}W^-_{\mu \nu})V^\nu\right.\\& +\left.  k_V W^+_\mu W^-_\nu V^{\mu \nu} +\frac{\lambda_V}{m_W^2}W_\mu^{\nu+}W_\nu^{-\rho}V_{\rho}^{\mu} +ig_4^VW_\mu^+W_\nu^-\right.\\&\left.(\partial^\mu V^\nu+\partial^\nu V^\mu) - ig_5^V\epsilon^{\mu \nu \rho \sigma}(W_\mu^+ \partial_\rho W_\nu^- - \partial_\rho W_\mu^+W_\nu^-)V_\sigma  
			\right.\\&+\left.\tilde{k}_VW_\mu^+W_\nu^-\tilde{V}^{\mu \nu} + \frac{\tilde{\lambda}_V}{m_W^2}W_\mu^{\nu+}W_\nu^{-\rho}\tilde{V}_\rho^{\mu}\right],
		\end{split}  	
	\end{equation}
	where $W^{\pm}_{\mu\nu} = \partial_\mu W^{\pm}_\nu-\partial_\nu W^\pm_\mu$, $V_{\mu\nu} = \partial_\mu V_\nu-\partial_\nu V_\mu$, $g_{WW\gamma} = -e$ and $g_{WWZ} = -e\cot\theta_W$, where $e$ and $\theta_W$ are the proton charge and weak mixing angle respectively. The dual field is defined as $\tilde{V}^{\mu\nu} = 1/2\epsilon^{\mu\nu\rho\sigma}V_{\rho\sigma}$, with Levi-Civita tensor $\epsilon^{\mu\nu\rho\sigma}$ follows a standard convention, $\epsilon^{0123}=1$. Within the SM, the couplings are given by $g_1^Z=g_1^\gamma = k_Z = k_\gamma = 1$ and all others are zero. The couplings $g_1^V,k^V$, and $\lambda^V$ are $CP$-even, while $g_4^V$ is odd in $C$ and $P$-even, $\widetilde{k}^V$ and $\widetilde{\lambda}^V$ are $C$-even and $P$-odd and the last coupling $g_5^V$ is $C$ and $P$-odd which make it $CP$-even. The search for anomalous signature based on parameterization of Eq.~(\ref{eqn:eff}) has some subtle issues as was pointed out in Ref.~\cite{Degrande:2012wf}; the theory can be expanded to infinitum by adding derivative normalized by mass of $W$ bosons and those terms are not suppressed at energies above $W$ mass, unless the anomalous couplings are very small. The other way to incorporate the anomalous contribution to $W^-W^+\gamma/Z$ while respecting the gauge structure of SM is known as effective field theory, or better known as SMEFT~\cite{Buchmuller:1985jz,Grzadkowski:2010es,Brivio:2017vri}. In this framework, the dimension-4 Lagrangian of SM is expanded by adding higher mass dimension terms constructed out of SM states, and each terms are normalized by some power of characteristic scale $\Lambda^{-1}$. Since, each terms are suppressed by $\Lambda^{-1}$, the effective Lagrangian can be truncated to some lowest order, and the subtle problem associated with Eq.~(\ref{eqn:eff}) are avoided. In presence of higher dimensional operators, the effective Lagrangian is written as~\cite{Buchmuller:1985jz,Hagiwara:1993ck},
	\begin{equation}
		\mathscr{L}_{\text{EFT}} = \mathscr{L}_{SM} + \frac{1}{\Lambda}\sum_i c_i^{(5)}\mathscr{O}_i^{(5)} + \frac{1}{\Lambda^2}\sum_i c_i^{(6)}\mathscr{O}^{(6)}_i + ...,
	\end{equation}
	where $c_i$ are the Wilson coefficient of higher order operators and index $i$ runs over a basis~$\{\mathscr{O}_i\}$. Considering the baryon-lepton number conservation, the operator with odd dimension can be neglected and $\Lambda$ being at the TeV range, we can safely truncate the above equation to some lowest order.
\begin{figure*}[!htb]
\centering
\begin{tabular}{ccc}
	\raisebox{1.5cm}{(a)} \makebox[0.32\textwidth][c]{\includegraphics[width=0.32\textwidth,height=3.5cm]{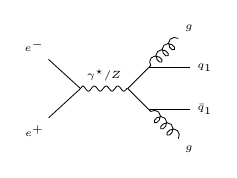}} &
	\raisebox{1.5cm}{(b)} \makebox[0.32\textwidth][c]{\includegraphics[width=0.32\textwidth,height=3.5cm]{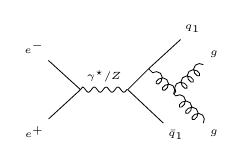}} &
	\raisebox{1.5cm}{(c)} \makebox[0.32\textwidth][c]{\includegraphics[width=0.32\textwidth,height=3.5cm]{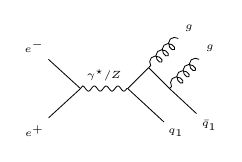}} \\
	\raisebox{1.5cm}{(d)} \makebox[0.32\textwidth][c]{\includegraphics[width=0.32\textwidth,height=3.5cm]{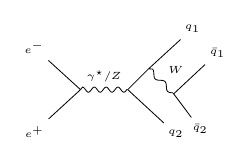}} &
	\raisebox{1.5cm}{(e)} \makebox[0.32\textwidth][c]{\includegraphics[width=0.32\textwidth,height=3.5cm]{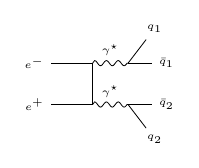}} &
	\raisebox{1.5cm}{(f)} \makebox[0.32\textwidth][c]{\includegraphics[width=0.32\textwidth,height=3.5cm]{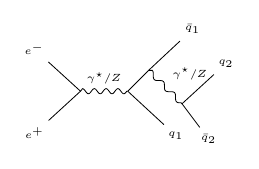}} \\
	\raisebox{1.5cm}{(g)} \makebox[0.32\textwidth][c]{\includegraphics[width=0.32\textwidth,height=3.5cm]{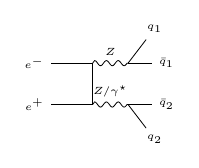}} &
	\raisebox{1.5cm}{(h)} \makebox[0.32\textwidth][c]{\includegraphics[width=0.32\textwidth,height=3.5cm]{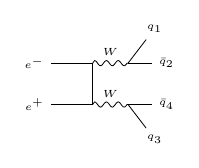}} &
	\raisebox{1.5cm}{(i)} \makebox[0.32\textwidth][c]{\includegraphics[width=0.32\textwidth,height=3.5cm]{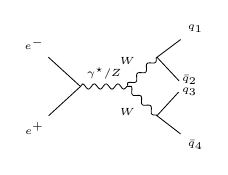}} \\
\end{tabular}
\caption{\label{fig:feyn}Schematic Feynman diagrams at lowest order for $4j$ events in final state at $e^-e^+$ collider. The top row amplitudes represents gluons in the final state, the second row contains zero-resonant amplitudes with four quarks in final states. The last row represents the one/two-resonant $ZZ/\gamma$, and $W^-W^+$ amplitudes.}
\end{figure*}
	In this article, we restrict ourself to dimension-6 operators, and the relevant operators are expressed in terms of dim-6 Lagrangian in HISZ~\cite{Hagiwara:1992eh, Hagiwara:1993ck,Degrande:2012wf} basis as,
	\begin{equation}
		\begin{aligned}
			\mathcal{L}^{(6)}&= \frac{c_{WWW}}{\Lambda^2}\text{Tr}[W_{\nu \rho}W^{\mu \nu} W_{\rho}^{\mu}] + \frac{c_W}{\Lambda^2} (D_\mu \Phi)^\dagger W^{\mu \nu} (D_\nu \Phi) \\
			&+ \frac{c_B}{\Lambda^2}(D_\mu \Phi)^\dagger B^{\mu \nu} (D_\nu \Phi) + \frac{c_{\widetilde{W}WW}}{\Lambda^2} \text{Tr}[\widetilde{W}_{\mu \nu} W^{\nu \rho} W^\mu_\rho] \\
			&+\frac{c_{\widetilde{W}}}{\Lambda^2}(D_\mu \Phi)^\dagger \widetilde{W}^{\mu \nu} (D_\nu \Phi),
		\end{aligned}
		\label{eqn:eftop}
	\end{equation}
	where $\Phi$ is the Higgs doublet and the covariant derivative and field tensors are defined as,
	\begin{equation*}
		\begin{aligned}
			D_\mu &= \partial_\mu  +\frac{i}{2}g\tau^i W_\mu^i + \frac{i}{2}g'B_\mu,\\
			W_{\mu \nu} &= \frac{i}{2}g\tau^i(\partial_\mu W_\nu^i-\partial_\nu W_\mu^i + g\epsilon_{ijk}W_\mu^i W_\nu^k),\\
			B_{\mu \nu} &= \frac{i}{2}g'(\partial_\mu B_\nu - \partial_\nu B_\mu).
		\end{aligned}
	\end{equation*}
	Here $g$, and $g^\prime$ are the $SU(2)_L$, and $U(1)_Y$ gauge couplings of SM. The first three operators in Eq.~(\ref{eqn:eftop}) are $CP$-even, and the last two are $CP$-odd.\\
	The presence of anomalous couplings at the vertex would modify both the angular and kinematic distributions of final decayed particles from the SM values. The effects of anomalous $W^-W^+\gamma/Z$ couplings on various kinematic and spin-related observables have been theoretically studied in several works~\cite{Buchalla:2013wpa, Choudhury:1999fz, Hagiwara:1992eh, Subba:2022czw, Subba:2023rpm, Rahaman:2019mnz, Bian:2015zha, Bian:2016umx, Tizchang:2020tqs, Rahaman:2019lab} and references therein. The most stringent experimental limits on these anomalous couplings are provided in Ref.~\cite{CMS:2021foa, CMS:2021icx}. In this study, we focus on the deviations in asymmetries related to the polarization and spin correlations of $W$ bosons in the $W^-W^+$ production process at an $e^-e^+$ collider at $\sqrt{s}=250$~GeV at leading order. The di-boson process are reconstructed in final four jets events using invariant mass and jet charge variable~(see Sec.\ref{sec:reconstruction}). The study of hadronic decay of $W$ bosons becomes important due to significant rate and the interference effects. At the leading order, the four jet final event topology in $e^-e^+$ collider results from three distinct sub-processes, and the details of each sub-processes are listed in Table~\ref{tab:subpro}. The representative Feynman diagrams for sub-processes listed in Table~\ref{tab:subpro} are shown in Fig.~\ref{fig:feyn}.
	\begin{table}[!h]
		\centering
		\caption{\label{tab:subpro}List of channels and the corresponding sub-processes and coupling order at the leading order for four jet topology in $e^-e^+$ collider.}
				\renewcommand{\arraystretch}{1.3}
		\begin{tabular*}{0.49\textwidth}
			{@{\extracolsep{\fill}}ccc@{}}\hline
			Channel&Sub-Process&Coupling order\\
			\hline
			{\tt CH1} &$e^-e^+\to q_1\bar{q}_1gg$&$\alpha_s^2\alpha_{EW}^2$\\
			{\tt CH2}&$e^-e^+\to q_1\bar{q}_2q_2\bar{q}_1$&$\alpha_{EW}^4$, $\alpha_s^2\alpha_{EW}^2$\\
			{\tt CH3}&$e^-e^+\to q_1\bar{q}_2 q_3 \bar{q}_4$&$\alpha_{EW}^4$\\
			\hline
		\end{tabular*}
	\end{table}
	We classify the production amplitudes in three channels based on the final states and the associated QCD and EW coupling order. The channel {\tt CH1} $\in \{(a),(b),(c)\}$ represent sub-process with gluon in the final state. The second channel {\tt CH2} is a mixture of both pure EW and QCD contributions, and all the amplitudes in middle and bottom row of Fig.~\ref{fig:feyn} are part of {\tt CH2}. The last type of channel, {\tt CH3} corresponds to a sub-process with quarks of different flavor at final state. The {\tt CH3} $\in \{(h),(i)\}$ are pure electroweak diagrams representing $W$ di-boson amplitudes, which are the signal sub-processes of our current analysis. One of the important feature that can be read off from this sort of classification is interference, and we are particularly interested in the interference of non-signal/{\tt Non-WW} with signal/{\tt WW} amplitudes. The three separate channels do not undergo interference due to different final state, but the {\tt CH2} becomes interesting channel as it contains both {\tt Non-WW} and {\tt WW} diagrams. The total matrix elements for four jet process can be written as,
	\begin{equation}
		\label{eq:int1}
		\mathcal{M}^2 = \mathcal{M}_{{\tt WW}}^2+2\mathcal{R}e\left(\mathcal{M}_{{\tt WW}}\mathcal{M}_{{\tt Non-WW}}\right)+\mathcal{M}_{{\tt Non-WW}}^2.
	\end{equation}
	\begin{figure*}[!htb]
		\centering
		\includegraphics[width=0.32\textwidth]{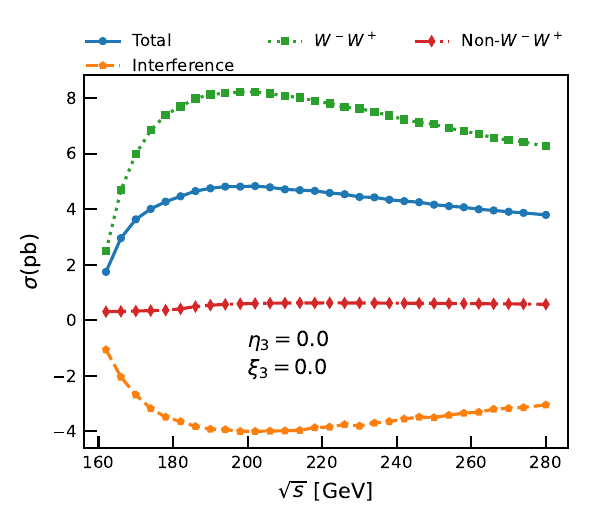}
		\includegraphics[width=0.32\textwidth]{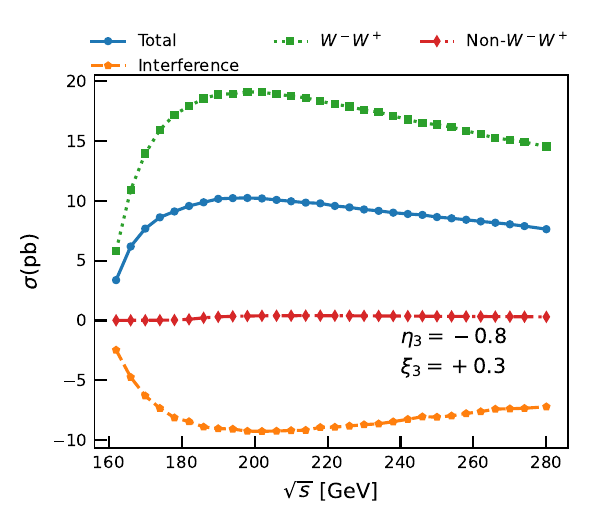}
		\includegraphics[width=0.32\textwidth]{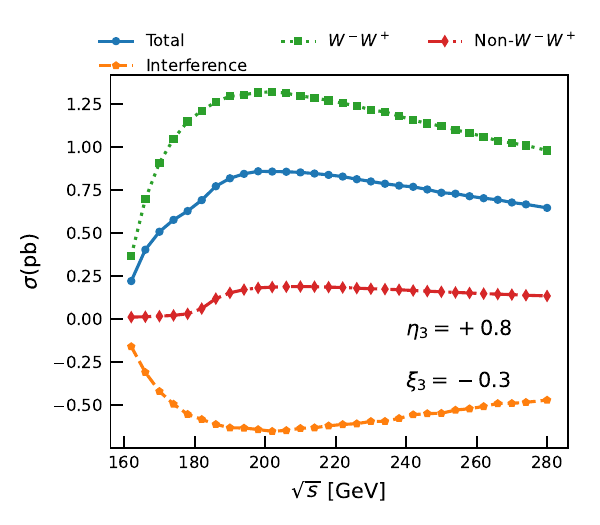}
		\caption{Cross~section as a function of center-of-mass energy for three set of events, i) signal~(green curve), ii) background~(red curve) and iii) signal plus background~(blue curve). The interference points~(orange curve) are calculated by using Eq.~(\ref{eq:int1}). The distribution shown for three set of initial beam polarization, Unpolarized case~(left panel), $(-0.8,+0.3)$~(middle panel), and the case with flipped polarization $(+0.8,-0.3)$ is shown in right panel.}
		\label{fig:smnopol}
	\end{figure*}
    The fractional contribution of the interference is at most given by $2|\mathcal{M}_{{\tt Non-WW}}|/|\mathcal{M}_{{\tt WW}}|$. Usually, in most of the studies, there is an incoherent mixture of two sets of diagrams, i.e., the signal and non-signal diagrams are studied separately. It is acceptable if the interference contribution is tiny like in semi-leptonic or leptonic decay channel of $WW$ di-boson process. However, many studies have been done that project significant kinematical changes due to interference~\cite{Ding:2019bkd,Baur:1989qt,RANFT1979122}, which can be highlighted here, in our four jet topology, due to large interference between $WW$ and zero-resonant amplitudes. The interference between different set of amplitudes can be highlighted in terms of relative phase difference, 
	\begin{equation}
		\label{eq:int2}
		\sigma \propto |\mathcal{M}_{{\tt WW}} + e^{i\phi}\cdot \mathcal{M}_{{\tt Non-WW}}|^2,
	\end{equation}
	where $\phi$ is the relative phase between two amplitudes. If we consider diagrams like $(g)$ of {\tt CH2} which are single resonant diagrams, they have minimal interference with {\tt WW} diagrams of {\tt CH2} as $\phi=\pi/2$. However, the phase difference between {\tt WW} and zero resonant amplitudes of {\tt CH2} becomes $\pi$ which leads to large negative interference. And since there are large number of such zero resonant amplitudes in {\tt CH2}, the interference would lead to significant reduction in overall rate.
	 \\
	To demonstrate the effect of interference on cross~section, we generate three sets of amplitudes, viz., $W^-W^+$-double resonant~$(\mathcal{M}_{{\tt WW}}^2)$, non-$W^-W^+$~$(\mathcal{M}_{{\tt Non-WW}}^2)$, and total~$(\mathcal{M}^2)$ using {\tt MadGraph5$\_$aMC$@$NLO~(v2.7.3)}~\cite{Alwall:2014hca}~({\tt MG5} henceforth). Then one can obtain the rate associated with interference using Eq.~(\ref{eq:int1}). 
	We depict the cross~sections for three different events and interference as a function of center-of-mass energy in Fig.~\ref{fig:smnopol}.
	We note that there is a large negative interference between the $W^-W^+$ and non-$W^-W^+$ amplitudes, which peaks around $\sqrt{s}\in \left[200,230\right]$~GeV. The peak rate of interference is $\approx$ half of the peak of $W^-W^+$ rate in case of unpolarized initial beam. Thus any result or analysis done without taking into account the coherent sum of the diagrams is off by a factor of around 2.5. This number becomes important once we delve into the precision study of new physics parameters, as the large interference provides a higher sensitivity to anomalous parameters.  
	\\
	The upcoming collider like ILC~\cite{Behnke:2013xla}, CLIC~\cite{Aicheler:2012bya,Linssen:2012hp,Lebrun:2012hj}, FCC-ee~\cite{FCC:2018evy}, and CEPC~\cite{CEPCStudyGroup:2018rmc,Li:2023kjt} are planned to collide polarized beams and in this article we consider the baseline setting of ILC with electron degree of polarization, $\xi_3=0.8$ and positron degree, $\eta_3 = 0.3$ along with the flipped polarization. The use of polarized beams could enhance the cross~section of the signal, thus increasing the overall significance, $\mathcal{S} = S/\sqrt{B}$ for $S>>B$, where $S$ and $B$ denotes the signal and background rate. In general the cross~section in presence of polarized beam~$(\eta_3,\xi_3)$ is defined as,
	\begin{equation*}
		\label{eqn:xsecpol}
		\begin{aligned}
			\sigma(\eta_3,\xi_3) &= \frac{1}{4}\left[(1+\eta_3)(1+\xi_3)\sigma_{\text{RR}} + (1-\eta_3)(1-\xi_3)\sigma_{\text{LL}}\right.\\&+\left. 
			(1+\eta_3)(1-\xi_3)\sigma_{\text{RL}} + (1-\eta_3)(1+\xi_3)\sigma_{\text{RL}}\right].   
		\end{aligned}     
	\end{equation*}
	Here $\sigma_{ij},i,j \in \{\text{R,L}\}$ represent the pure chiral cross~sections. To understand the impact of beam polarization on the cross-section, we simulate three distinct sets of events: pure signal, background, and interference at the SM point. The pure signal corresponds to the amplitudes represented by panels $(h)$ and $(i)$ of Fig.~\ref{fig:feyn}. In contrast, the background encompasses the class of amplitudes illustrated by all panels except $(h)$ and $(i)$ of Fig.~\ref{fig:feyn}.
	\\ 
	The beam polarization, $(\eta_3,\xi_3) = (-0.8,+0.3)$, enhances the cross~section for signal, and consequently, the rate of interference in the negative direction while the rate for flipped polarization is not significantly increased (see Fig.~\ref{fig:smnopol} middle and right panel). The total rate thus gets lowered once we take into account the effect of interference in the analysis. The important point is that a large contribution to the total rate in both polarization sets comes from interference and it is properly included in rest of our study in this paper.
	\\\\
	The plan of the paper is as follows, in Section~\ref{sec:obs}, we discuss polarization and spin correlation asymmetries which will be employed in this article to constrain anomalous couplings. We also highlight the fact that some of these parameters requires flavor tagging of final jets. In Section~\ref{sec:reconstruction}, we discuss the methodology to reconstruct $W^-W^+$ topology, and tag charged $W$ boson using jet variable. The flavor tagging of final jets achieved with boosted decision trees is discussed. Section~\ref{sec:probe} discusses the methodology and the obtained limits on anomalous couplings. We finally conclude in Section~\ref{sec:conclude}.
	\section{Angular distribution}
	\label{sec:obs}
	The spin of a particle is a fundamental property that influences the Lorentz structure of its interactions with other SM particles, and understanding these interactions is essential for probing SM or any beyond SM theories~\cite{Boudjema:2009fz,Aguilar-Saavedra:2017zkn,Aguilar-Saavedra:2015yza}. On the other hand, the spin of a particle also dictates the angular distribution of decayed particles, which can be influenced by the change in the interaction. Thus, the change in couplings due to the presence of new physics may lead to deviation of various angular functions of final state particles from the SM value. Based on the value of spin of a decaying particle, we can quantify these angular functions in terms of polarization parameters. In particular, for a spin-1 particle like $W$ boson, 8 independent polarization parameters holds the information of production dynamics which can be obtained by using the angular distribution of the decayed daughter. And in a process where two spin-1 boson~($W^-W^+$) are produced, the production dynamics are encoded in $16$ polarizations, and $64$ spin correlations, which can be obtained in terms of the polar and azimuth angle of final decayed fermions. We list the correlators $C_i^\prime$s in terms of angular functions of final decayed fermions, and the related generators in Table~\ref{tab:corr}.
	\begin{table}[!htb]
		\centering
		\caption{\label{tab:corr}List of angular functions that are associated with the distribution of final decayed fermions and the corresponding generators.}
		\renewcommand{\arraystretch}{1.5}
		\begin{tabular}{ccc}
			\hline
			Correlators & Functions & Generators \\
			\hline
			$C_1$ & $1$ & $J_1=\mathbb{I}$\\
			$C_2$ & $\sin\theta\sin\phi$ & $J_2 = S_x/2$ \\
			$C_3$ & $\sin\theta\cos\phi$ & $J_3 = S_y/2$ \\
			$C_4$ & $\cos\theta$ & $J_4 = S_z/2$ \\
			$C_5$ & $C_2\cdot C_3$ & $J_5 = \left(S_xS_y + S_yS_x\right)/2$\\ 
			$C_6$ & $C_2\cdot C_4$ & $J_6 = \left(S_xS_z+S_zS_x\right)/2$\\
			$C_7$ & $C_3\cdot C_4$ & $J_7 = \left(S_yS_z+S_zS_y\right)/2$\\
			$C_8$ & $C_4^2- C_5^2$ & $J_8 = \left(S_xS_x - S_yS_y\right)/2$\\
			$C_9$ & $\sqrt{1-C_4^2}\left(3-4(1-C_4^2)\right)$&$J_9=\sqrt{3}\left(S_zS_z/2-\mathbb{I}/3\right)$\\
			\hline 
		\end{tabular}
	\end{table}
	\\
	The $S_i, i \in \{x,y,z\}$ are the three spin-1 operators, and the $J_i^\prime$s matrices have orthonormal properties $\text{Tr}\left[J_iJ_j\right] = \delta_{ij}/2$~(3 if $i = j = 1)$. For a process where two spin-1 $W$ boson decays hadronically, the joint angular distribution of the final decayed fermions are written as~\cite{Rahaman:2021fcz},
	\begin{equation}
		\label{eqn:jdm}
		\begin{aligned}
				&\frac{1}{\sigma}\frac{d\sigma}{d\Omega_{j_1}d\Omega_{j_2}} = \sum_{\lambda_{W^-},\lambda_{W^-}^\prime,\lambda_{W^+},\lambda_{W^+}^\prime}\\&\rho_{W^-W^+}\left(\lambda_{W^-},\lambda_{W^-}^\prime,\lambda_{W^+},\lambda_{W^+}^\prime\right)\\&\Gamma_{W^-}\left(\lambda_{W^-},\lambda_{W^-}^\prime\right)\Gamma_{W^+}\left(\lambda_{W^+},\lambda_{W^+}^\prime\right),			
		\end{aligned}		
	\end{equation} 
	where the production density matrix $\rho_{W^-W^+}$ can be written in basis space of $9\times 9$ matrices formed by the tensor product $\{\mathbb{I}\otimes \mathbb{I},\mathbb{I}\otimes J_i,J_i \otimes \mathbb{I}, J_i\otimes J_j \}$. Then, the density matrix is given by (see for example~\cite{Rahaman:2021fcz}),
	\begin{equation}
		\label{eqn:rho}
		\begin{aligned}
			\rho_{W^-W^+} &= \mathbb{I}\otimes \mathbb{I} + \sum_{i=2}^4\left(P_i^{(2)} \mathbb{I}\otimes J_i^{(2)} + P_i^{(1)}J_i^{(1)}\otimes \mathbb{I}\right) \\&+ \sum_{i=2}^9C_{ij}^{(12)}J_i^{(1)}\otimes J_j^{(2)}.
		\end{aligned}	
	\end{equation}
	Here, $P^\prime$s are eight independent polarization, and $C_{ij}^{(12)}$ are $64$ correlation  parameters. These parameters can be obtained from the asymmetries in the correlators as,
	\begin{equation}
		A_{ij} = \frac{\sigma(C_i^{(1)}C_j^{(2)} > 0) - \sigma(C_i^{(1)}C_j^{(2)}<0)}{\sigma(C_i^{(1)}C_j^{(2)} > 0) + \sigma(C_i^{(1)}C_j^{(2)}<0)},
	\end{equation}
	where $A_{i0},A_{0j}, i,j\in \{1,..,8\}$ are the polarization asymmetries, and $A_{ij}, i,j \in \{1,..,8\}$ are the spin correlation asymmetries. One can obtain the exact relation between the asymmetries with polarization, and spin correlation parameters by doing a partial integration of Eq.~(\ref{eqn:jdm})~(see Ref.~\cite{Rahaman:2021fcz}). In Eq.~(\ref{eqn:jdm}), the $\Gamma$ matrices are the decay density matrix of $W$ boson and is given in Appendix~\ref{sec:sdm}. 
	\begin{figure}
		\centering
		\includegraphics[width=0.49\textwidth]{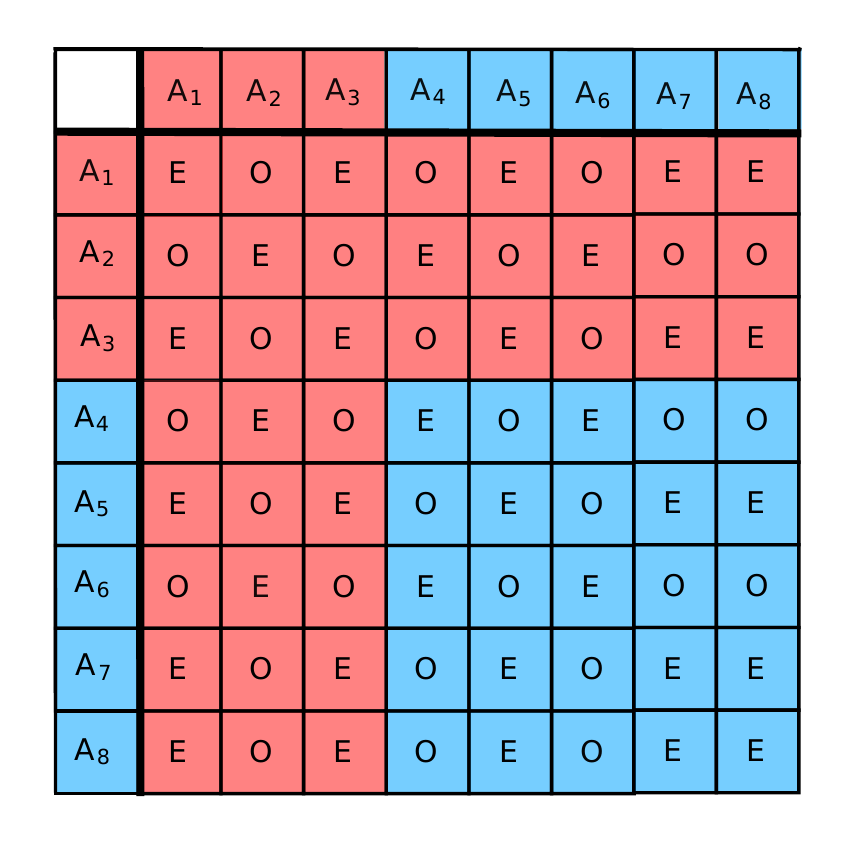}
		\caption{\label{fig:cp}Representative diagram showing the list of all polarization and spin-correlation asymmetries along with their $CP$ structure and flavor dependency.}
	\end{figure}
	The correlators $C_{ij}$ have different property under $CP$ transformation, for e.g. the correlators $C_1$ is $CP$-even, and $C_2$ is $CP$-odd. On top of that, the construction of asymmetries associated with the vector polarizations, vector-vector, and vector-tensor correlations requires decay products of $W^\prime$s to be tagged. Since, we are considering an events satisfying a $W^-W^+$ topology, the final daughter of boson can be either \emph{up}-type or \emph{down}-type jets. We leverage the classification technique of boosted decision trees~(BDT) algorithm to flavor tag the final jets. The complete $CP$ structure along with flavor dependence of $A_{ij}$ are listed in Fig.~\ref{fig:cp}. The asymmetries related to $CP$-even functions are denoted as "E", and those which are $CP$-odd is denoted by "O". And the flavor dependence asymmetries are denoted as grid with light red color, while those which do not require a flavor identification are denoted in grid with light blue color. In total there are $44$ $CP$-even, and $36$ $CP$-odd functions, $45$ flavor dependent, and $35$ flavor independent functions in the case of two spin-1 $W^-W^+$ production process. The methodology to select the required event topology and flavor tag the final jets are discussed in next section.
	
	\section{Reconstruction of $W$ boson and Jet Flavor Tagging}
	\label{sec:reconstruction}
	\begin{figure*}[!htb]
		\centering
		\includegraphics[width=0.49\textwidth]{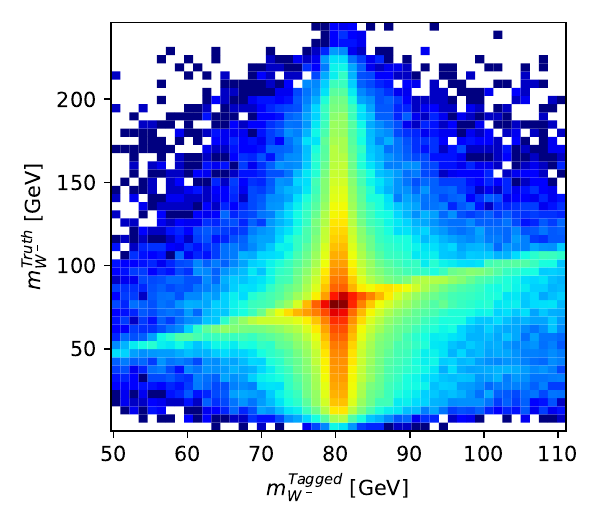}
		\includegraphics[width=0.49\textwidth]{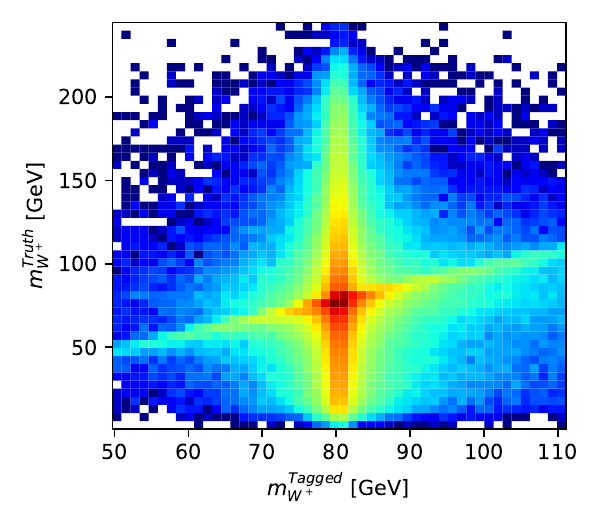}
		\caption{\label{fig:reocmass}Normalized distribution of true vs. tagged $W$ boson mass in the $e^-e^+\to 4j$ process at $\sqrt{s}=250$~GeV. The reconstruction of two $W$ boson are described in Sec.~\ref{sec:reconstruction}.}
	\end{figure*}
	This section discusses the methodology used to reconstruct the $W$ boson from four final jets in a simulated collision experiment. The generation of events at leading order is done using {\tt MG5}, and {\tt Pythia8} is used for showering and hadronization and finally the clustering are done using {\tt FastJet}. All final state visible particles with transverse momentum $p_T > 0.3$ GeV are considered for clustering. The particles are clustered using the anti-$k_T$ algorithm with a jet radius of $R = 0.7$. The jets obtained from the anti-$k_T$ clustering are then used as inputs for a second phase of clustering, which is achieved using the $k_T$ clustering algorithm with a $R = 1.0$ jet radius. The use of the $k_T$ algorithm allows for the merging of soft jets with hard ones. We select the four hardest jets for further analysis.
	\begin{figure*}[!htb]
		\centering
		\includegraphics[width=0.49\textwidth,height=7cm]{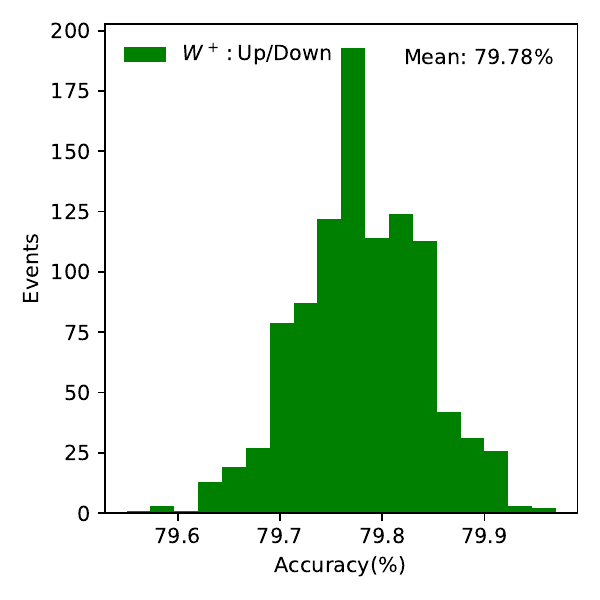}
		\includegraphics[width=0.49\textwidth,height=7cm]{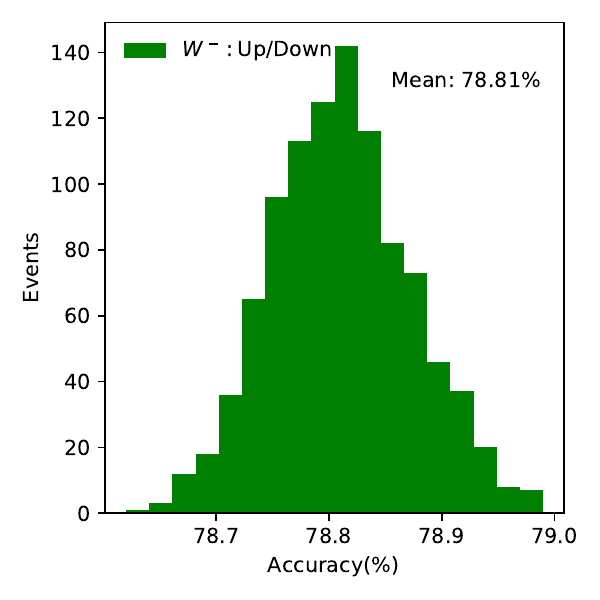}
		\caption{Accuracy on classifying the two daughter jets of tagged $W$ boson as up-type or down-type.}
		\label{fig:xgbacc}
	\end{figure*}
	We consider combining the jets in pairs to mimic the daughter jets of the $W$ boson. If only the signal diagrams were considered, the geometric distance $\Delta R = \sqrt{\Delta\eta^2 + \Delta\phi^2}$ could be used to tag the daughter jets of individual $W$ bosons. However, in our case, the final events are the result of all possible Feynman diagrams in the leading order, including interference between signal and non-signal diagrams. This makes the separation of signal and background non-trivial. We reconstruct the $W$ boson as follows,
	\begin{itemize}
		\item Assume the four jets in an events are $J \in \{J1,J2,J3,J4\}$. There exist three unique combinations to make a pair.
		\item For each pair we calculate jet charge,~$Q_J^k$~\cite{Field:1977fa,Waalewijn:2012sv,Chang:2013rca,Kang:2023ptt} and is calculated as,
		\begin{equation}
			\label{eqn:jetcharge}
			Q_J^k = \frac{1}{p_{TJ}^k}\sum_iQ^ip_{Ti}^k,
		\end{equation}
		where $p_T^J$ is the transverse momenta of the jet and $p_T^i,Q^i$ are transverse momenta and charge of each particle inside the jet and $k$ is some real number $\mathbb{R}\in\left[0,1\right]$. For our case, we find $k=0.2$ produces statistically better resolution between two pairs.
		\item Of three possible pairs, we only keep those pair sets for which the jet charge product is negative.
		\item If the event contains more than one set of pair representing opposite sign $W^\prime$s boson, we select the one with minimum sum in $W$ boson mass deviation, $$||m_{Ji} - m_W| + |m_{Jj} - m_W||$$.
		\item For the surviving pair of jets, we tagged the jet with the negative jet charge to be $W^-$ boson and the jet with the positive jet charge to be $W^+$ boson.
	\end{itemize}
	We note that approximately $85\%$ of total events survived all the above-listed selection criteria. The 2-d normalized distribution for true vs. tagged $W$ boson mass is shown in Fig.~\ref{fig:reocmass}, where the two $W$ boson are tagged using the above described algorithm. We note that the mass of majority of reconstructed boson are within the deviation of true $W$ mass.

	Next we used the constituent jets of the reconstructed $W$ boson for flavor tagging. Once two pair of jets are tagged as $W$ bosons, we further used the information of this combined jet to flavor tag the final daughter. The necessity to identify the flavor of final jets comes from the observation that some of the polarization and related spin correlation asymmetries average out unless the flavor identity is known, and the overall sensitivity to anomalous couplings decreases. The list of asymmetries in terms of flavor dependent and independent are shown in Fig.~\ref{fig:cp}. We developed a boosted decision tree ~(BDT) to tag the final jets of $W$ boson as initiated by \emph{up/down}-type quarks. The input to BDT models is obtained from the constituents of jets. The network is trained using a labeled data for \emph{up/down}-type jets from the $W^-W^+$ resonant events. The truth labeling is done using the distance $\Delta R = \sqrt{\Delta\phi_{qj}^2 + \Delta\eta_{ej}^2}$ between the initial quark and the final four jets. The features used for the tagging purpose are similar to that describe in Ref.~\cite{Subba:2023rpm}, and are listed in Appendix~\ref{sec:app2} for completeness. On top of the features listed in~\cite{Subba:2023rpm}, we also constructed the jet charge for each jet for different values of $k \in \{0.2,0.4,0.6,0.8\}$ using Eq.~(\ref{eqn:jetcharge}).
	The BDT is  implemented in {\tt XGBoost} for binary classification with the following parameters,
	\begin{itemize}
		\item The sub-sample ratio of columns when constructing each tree, {\tt colsample$\_$bytree} = 0.8,
		\item Step size shrinkage used in the update to prevent overfitting, {\tt eta} = 0.3,
		\item Minimum loss reduction required to make a further partition on a leaf node of the tree, {\tt Gamma} = 1.5,
		\item Maximum depth of a tree, {\tt max$\_$depth} = 5,
		\item Number of trees, {\tt n$\_$estimators} = 300,
		\item L1 regularization term on weights, {\tt alpha} = 1.5,
		\item L2 regularization term on weights, {\tt lambda} = 1.5.
	\end{itemize}
	One million datasets were used for training, and $5\times 10^5$ events were used for testing. The pre-processing of the data is done by scaling to unit variance, $z = (x - \bar{x})/s$, where $\bar{x}$, and $s$ are the mean and standard deviation of datasets.
	To obtain the robust efficiency of our models, we select random $60\%$ of test data and calculate the accuracy on sub-sample. We iterate this $1000$ times with different sub-samples in each iteration. The distribution of accuracy obtained is shown in Fig.~\ref{fig:xgbacc}, and we note that the average accuracy for classification of two jets as \emph{up/down}-type in case of tagged $W^+$ and $W^-$ are $79.78\%$ and $78.81\%$, respectively. Though the network is trained only using the unpolarized datasets, it has been reported in Ref.~\cite{Subba:2023rpm} that the tagger is blind to the initial beam polarization and any network can be used for cross tagging. Thus, we used the unpolarized tagger for further jet tagging with beam polarization. The tagged jets are used to construct the polarization and spin-correlation asymmetries of tagged $W$ bosons, which are further employed to constrain the anomalous couplings.\\
	\section{Probe of anomalous couplings}
	\label{sec:probe}
		\begin{table*}[!htb]
		\centering
		\caption{\label{tab:Tab1}Experimental and theoretical~(semi-leptonic and full-hadronic channel) limits on anomalous couplings~$c_i$~(TeV$^{-2}$) at $95\%$ confidence level obtained by varying one parameter at a time. The theoretical limits are listed for $\sqrt{s}=250$~GeV, $\mathcal{L}=100$~fb$^{-1}$ and beam polarization~$(\eta_3,\xi_3) = (\pm0.8,\mp0.3)$.\\ }
		\renewcommand{\arraystretch}{1.5}
		\begin{ruledtabular}
			\begin{tabular}{lccc}
				Parameters~$c_i$& Experimental~(TeV$^{-2}$)&Semi-Leptonic~(TeV$^{-2}$)~\cite{Subba:2023rpm}&Full-Hadronic~(TeV$^{-2}$) \\ \hline
				$c_{WWW}/\Lambda^2$&$[-0.90,+0.91]$ CMS~\cite{CMS:2021foa}&$[-0.92,+0.92]$&$\left[-0.27,+0.25\right]$\\
				$c_W/\Lambda^2$&$[-2.10,+0.30]$ CMS~\cite{CMS:2021icx}&$[-0.67,+0.67]$&$\left[-0.07,+0.07\right]$\\
				$c_B/\Lambda^2$&$[-8.78,+8.54]$ CMS~\cite{CMS:2021foa}&$[-1.46,+1.46]$&$\left[-0.34,+0.34\right]$\\
				$c_{\widetilde{W}}/\Lambda^2$&$[-20.0,+20.0]$ CMS~\cite{CMS:2021foa}&$[-4.62,+4.62]$&$\left[-0.70,+0.70\right]$\\
				$c_{\widetilde{WWW}}/\Lambda^2$&$[-0.45,+0.45]$ CMS~\cite{CMS:2021foa}&$[-1.00,+1.00]$&$\left[-0.33,+0.33\right]$ \\
			\end{tabular}
		\end{ruledtabular}
	\end{table*}
	This section discusses the methodology employed to constrain the anomalous couplings $c_i$. As discussed in above section, $16$ polarization and $64$ spin-correlation asymmetries exist in the case of pair produced $W$ boson. In order to effectively constrain the anomalous couplings, we have categorized all observables into eight distinct intervals of $\cos\theta^{W^-}$, where $\theta^{W^-}$ represents the production angle of the $W^-$ boson in the laboratory frame. Within each bin, we have identified $1$ cross-section, $16$ polarizations, and $648$ spin-correlations asymmetries, resulting in a total of $648$ observables. These observables have been computed for both the SM and several sets of benchmark anomalous points. For each bin of SM and anomalous points, we construct different $648$ observables, and these values are used for numerical fitting to obtain semi-analytical relation between those observables and anomalous couplings. For cross~section, which is a $CP$-even observable, the following parametric function is used,
	\begin{equation}
		\begin{split}
			\sigma(\{c_i\}) &= \sigma_0 + \sum_{i=1}^3\sigma_ic_i + \sum_{j=1}^5\sigma_{jj}c_j^2 + \sum_{i> j}^3\sigma_{ij}c_ic_j \\&+ \sigma_{45}c_4c_5,
		\end{split}
		\label{eqn:cpeven}
	\end{equation}
	where couplings $c_i\in \{c_1,c_2,c_3\}$ corresponds to $CP$-even and $c_i\in\{c_4,c_5\}$ corresponds to $CP$-odd couplings. In the case of asymmetries, the denominator is cross~section, while the numerator corresponds to difference in cross~section. The numerator~($\mathcal{A} = \
	\mathcal{A} \times \Delta\sigma$) of the $CP$-odd asymmetries are fitted using function,
	\begin{equation}
		\Delta\sigma(\{c_i\}) = \sum_{i=4}^5\sigma_ic_i + \sum_{i=1}\sigma_{i4}c_ic_4 + \sum_{i=1}\sigma_{i5}c_ic_5.  
		\label{eqn:cpodd}
	\end{equation}
	\begin{figure*}[!htb]
		\centering
		\includegraphics[width=0.49\textwidth]{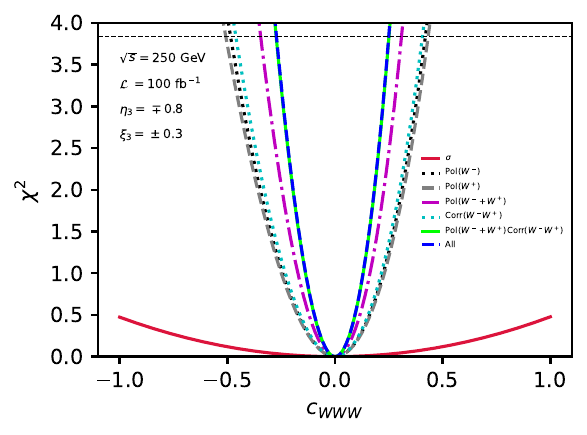}
		\includegraphics[width=0.49\textwidth]{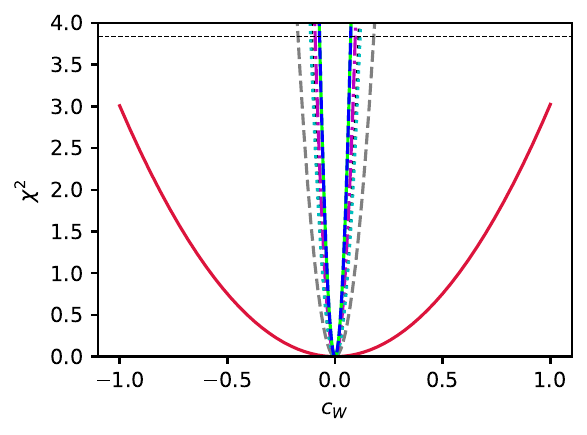}
		\includegraphics[width=0.49\textwidth]{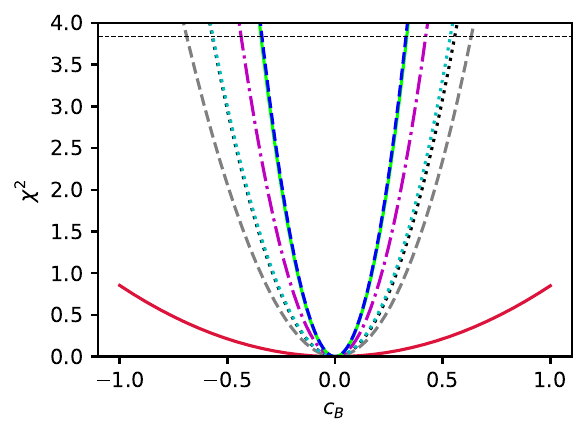}
		\includegraphics[width=0.49\textwidth]{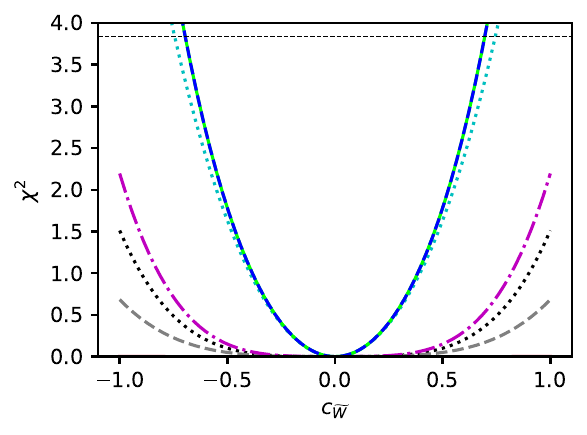}
		\includegraphics[width=0.49\textwidth]{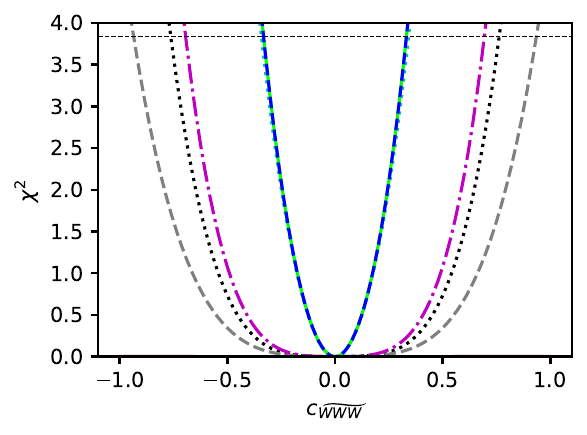}
		\caption{Chi-squared distribution for cross~section~($\sigma$), polarization of two $W$ boson~({\tt Pol($W^-$),Pol($W^+$)}), spin correlation~({\tt Corr$(W^-W^+)$}) and their combinations as a function of one anomalous couplings at a time. The analysis is done at $\sqrt{s}=250$~GeV, $\mathcal{L}=100$~fb$^{-1}$ with two set of beam polarization, $(\eta_3,\xi_3) = (\mp0.8,+\pm0.3)$, and zero systematic errors. The horizontal line at $\chi^2=3.84$ represent limit on anomalous couplings at $95\%$ CL.}
		\label{fig:chione}
	\end{figure*}
	Next, we study the sensitivity of cross~section, polarization, and spin correlation asymmetries by constructing the $\chi^2$ as a function of anomalous couplings. For an observable $\mathcal{O}$, we find the chi-squared distance between the SM and SM plus anomalous point in the presence of two sets of beam polarization, $(\mp\eta_3,\pm\xi_3)$ as,
	\begin{equation}
		\label{eqn:chisq}
		\begin{split}
			&\chi^2\left(\mathcal{O},c,\pm\eta_3,\mp\xi_3\right) = \\&\sum_{i,j}\left[\left(\frac{\mathcal{O}^i_j(c,+\eta_3,-\xi_3)-\mathcal{O}^i_j(0,+\eta_3,-\xi_3)}{\delta\mathcal{O}^i_j(0,+\eta_3,-\xi_3)}\right)^2 \right.\\&+\left. \left(\frac{\mathcal{O}^i_j(c,-\eta_3,+\xi_3)-\mathcal{O}^i_j(0,-\eta_3,+\xi_3)}{\delta\mathcal{O}^i_j(0,-\eta_3,+\xi_3)}\right)^2  \right], 
		\end{split}
	\end{equation}
	where indices $i,j$ represent observables and bins, respectively. The $\delta\mathcal{O}$ is the estimated error on observable $\mathcal{O}$, for cross~section it is,
	\begin{equation}
		\delta\sigma = \sqrt{\frac{\sigma}{\mathcal{L}} + (\epsilon_\sigma\sigma)^2},
	\end{equation}
	and for various asymmetries, the error is given by
	\begin{equation}
		\delta\mathcal{A} = \sqrt{\frac{1-\mathcal{A}^2}{\mathcal{L}\sigma}+\epsilon_A^2}.
	\end{equation}
	Here $\epsilon_\sigma$ and $\epsilon_\mathcal{A}$ are the fractional systematic error in cross~section~$\sigma$ and asymmetries~$\mathcal{A}$, respectively and $\sigma$ and $\mathcal{L}$ are the SM cross~section and integrated luminosity. The analysis is done for different values of integrated luminosity,
	\begin{equation}
		\mathcal{L} \in \{100~\text{fb}^{-1},250~\text{fb}^{-1},1000~\text{fb}^{-1},3000~\text{fb}^{-1}\},
		\label{eqn:lumi}
	\end{equation}
	where we have used $\mathcal{L}/2$ for each set of beam polarization and systematic errors,
	\begin{equation}
		(\epsilon_\sigma,\epsilon_\mathcal{A}) \in \{(0.0,0.0),(0.5\%,0.25\%),(2\%,1\%)\}.
		\label{eqn:syst}
	\end{equation}
	\begin{figure*}[!htb]
		\centering
		\includegraphics[width=0.32\textwidth]{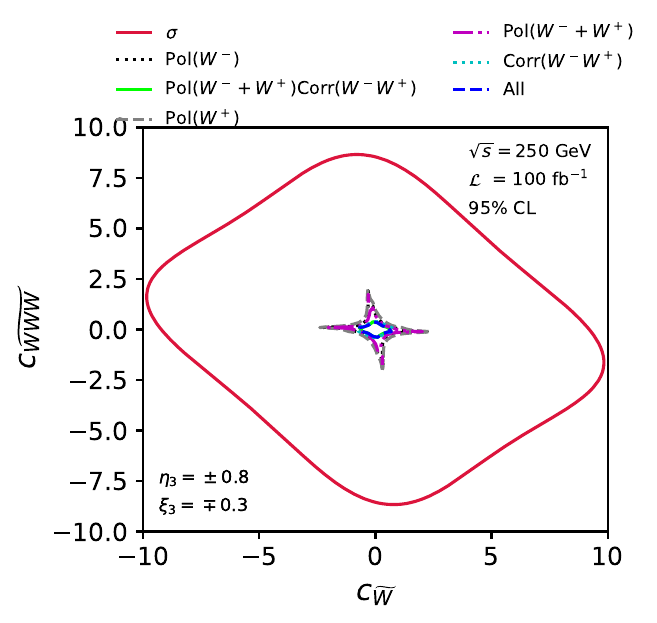}
		\includegraphics[width=0.32\textwidth]{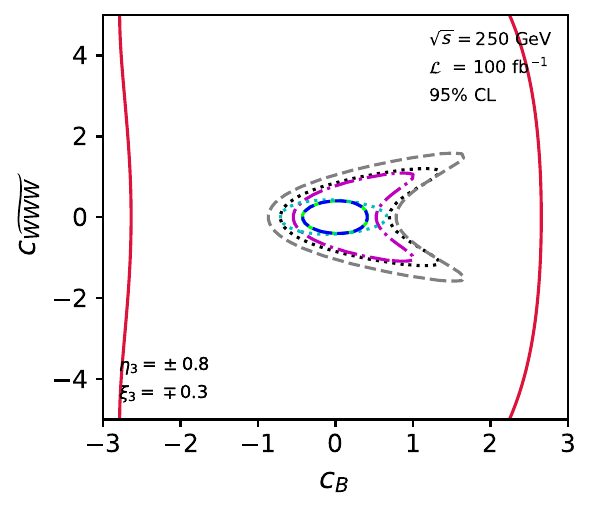}
		\includegraphics[width=0.32\textwidth]{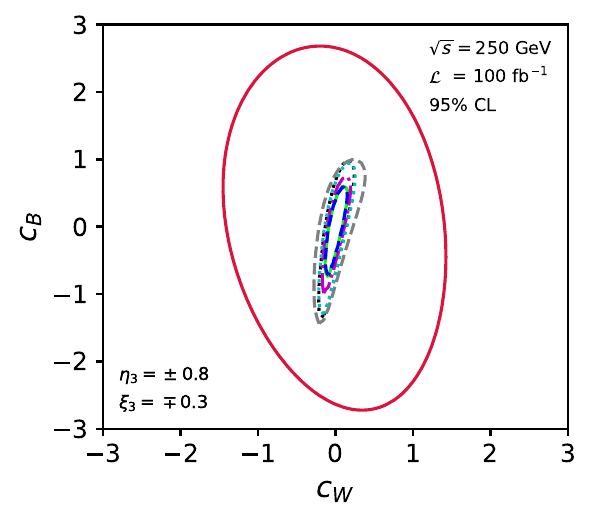}
		\caption{Two dimensional $95\%$ CL contours computed with $\chi^2=5.996$ function for cross~section, polarization, and spin correlation asymmetries as a function of two anomalous couplings at a time. The $\chi^2$ is computed at $\sqrt{s}=250$~GeV, $\mathcal{L}=100$~fb$^{-1}$ and two set of beam polarization $(\mp0.8,\pm0.3)$.}
		\label{fig:twodchi}
	\end{figure*}
	We conduct a sensitivity study to investigate the impact of anomalous couplings on various sets of observables using Eq.~(\ref{eqn:chisq}). The observables studied are the cross~section~($\sigma$), polarizations of two tagged $W$ bosons~(Pol($W^-$), Pol($W^+$)), spin correlation~(Corr($W^-W^+$)), a combined polarization~(Pol($W^-+W^+$)), a combined polarization and spin-correlation~(Pol($W^-+W^+$)Corr($W^-W^+$)), and a combination of all observables~(All). The sensitivity of these different sets of observables is depicted in Fig.~\ref{fig:chione} as a function of one anomalous coupling while keeping the other couplings to zero.\\
	From the analysis, we observe that the contribution of the cross-section (represented by the red curve) is minimal for both $CP$-even couplings and $CP$-odd couplings, compared to the spin-related observables. It is because in presence of $CP$-odd couplings, the contribution to cross~section only comes at $1/\Lambda^4$ order whereas the $CP$-even couplings can contribute to cross~section at both $1/\Lambda^2$ and $1/\Lambda^4$ terms. We list the $95\%$ confidence level~(CL) one parameter limits on anomalous couplings $c_i$ in Table.~\ref{tab:Tab1}, where we have listed the experimental and limits from semi-leptonic channel~\cite{Subba:2023rpm} for comparison.

	The $95\%$ CL one parameter limits for all the anomalous couplings obtained at the full-hadronic channel are tighter than the experimental counterpart obtained at CMS. The limits on $CP$-even couplings, $c_i\in\{c_{WWW},c_B\}$, are tighter by a factor of $3.1,25.8$, respectively than the corresponding experimental limits. For $c_W$, the upper and lower bounds obtained from full-hadronic channel is tighter by a factor of $30,$ and $4.3$, respectively. In the case of $CP$-odd couplings, bounds on $c_{\widetilde{W}},$ and $c_{\widetilde{WWW}}$ are tighter by a factor of $28.6$, and $1.4$, respectively than the experimental bounds. The sensitivity of $c_{\widetilde{W}WW}$ to cross~section directly depends on the partonic center of mass energy, which have been exploited by LHC leading to tighter experimental constraint as compare to other couplings. Whereas, in the lower $\sqrt{s}$ at $e^-e^+$ collider, it has been complemented by the large number of sensitive asymmetries observables. In comparison to the semi-leptonic channel, the limits on $c_i\in \{c_{WWW},c_W,c_B,c_{\widetilde{W}},c_{\widetilde{W}WW}$\} set from full-hadronic channel are tighter by a factor of $3.4,9.6,4.3,6.6,3.0$, respectively. 
	\\	
	Next, we study the sensitivity of different sets of observables as a function of two anomalous couplings at a time while keeping others to zero. For graphical representation, we present $\chi^2=5.996$ contours for different sets of observables and different combinations of anomalous couplings~$\{(c_{\widetilde{W}},c_{\widetilde{W}WW}),(c_B,c_{\widetilde{W}WW}),(c_W,c_B)\}$ in Fig.~\ref{fig:twodchi}. 
	In the case when both the anomalous couplings are $CP$-odd, the cross~section provides the poorest limits due to the negligible contribution from the $1/\Lambda^4$ term, and the final bounds are dominated by spin-related observables. While in the case when one of the parameters is $CP$-even, and the other is $CP$-odd, limits due to cross~section are tighter on the x-axis, i.e, $c_B$ and loose on $c_{\widetilde{WWW}}$. Since there may exist a cancellation of the linear term in the case when both the parameters are $CP$-even, we note a tighter limit on $\sigma_i = \sigma_{j}\frac{c_{j}}{c_i}$ axis and weaker limits on orthogonal axis.
	\\
	\begin{figure*}[!htb]
		\centering
		\includegraphics[width=0.32\textwidth,height=6cm]{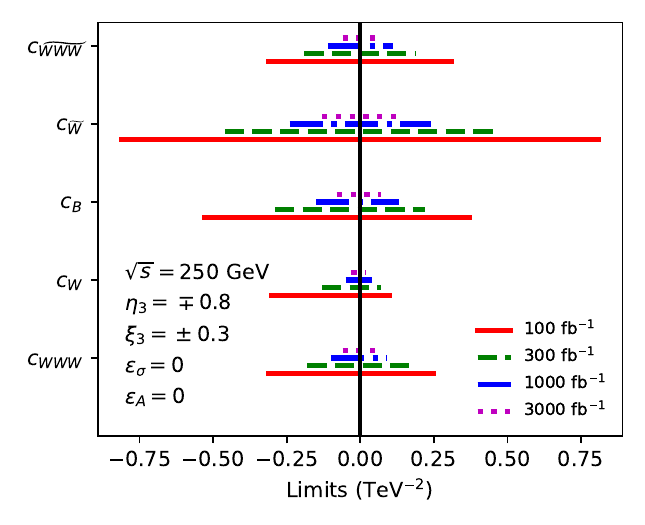}
		\includegraphics[width=0.32\textwidth,height=6cm]{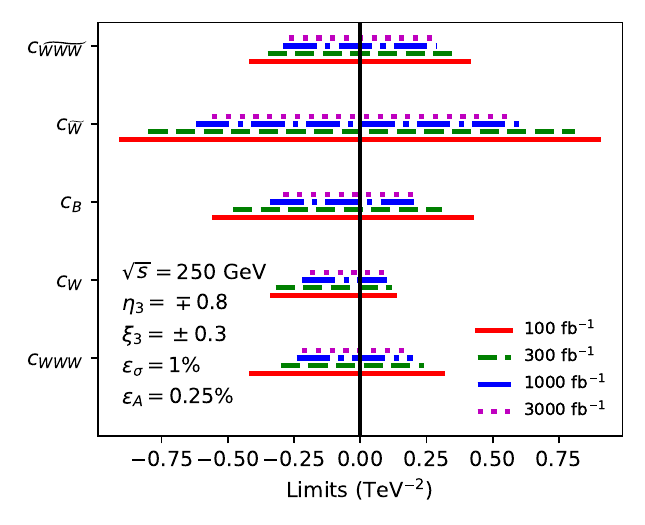}
		\includegraphics[width=0.32\textwidth,height=6cm]{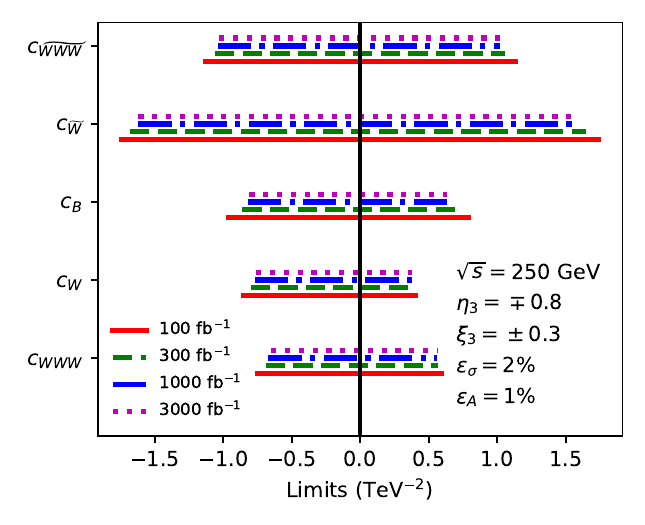}
		\caption{The graphical visualizations of 95$\%$ BCI limits obtained from MCMC global fits on the anomalous couplings $c_i$ for a different set of systematic error, $(\epsilon_\sigma,\epsilon_{A})$ = (0,0) in the leftmost panel, ($1\%,0.25\%$) in the middle panel and ($2\%,1\%$) in the right-most panel. The limits are obtained at $\sqrt{s}=250$ GeV and luminosity given in Eq.~(\ref{eqn:lumi}).}
		\label{fig:one95}
	\end{figure*}
	\begin{figure*}[!htb]
		\centering
		\includegraphics[width=0.32\textwidth]{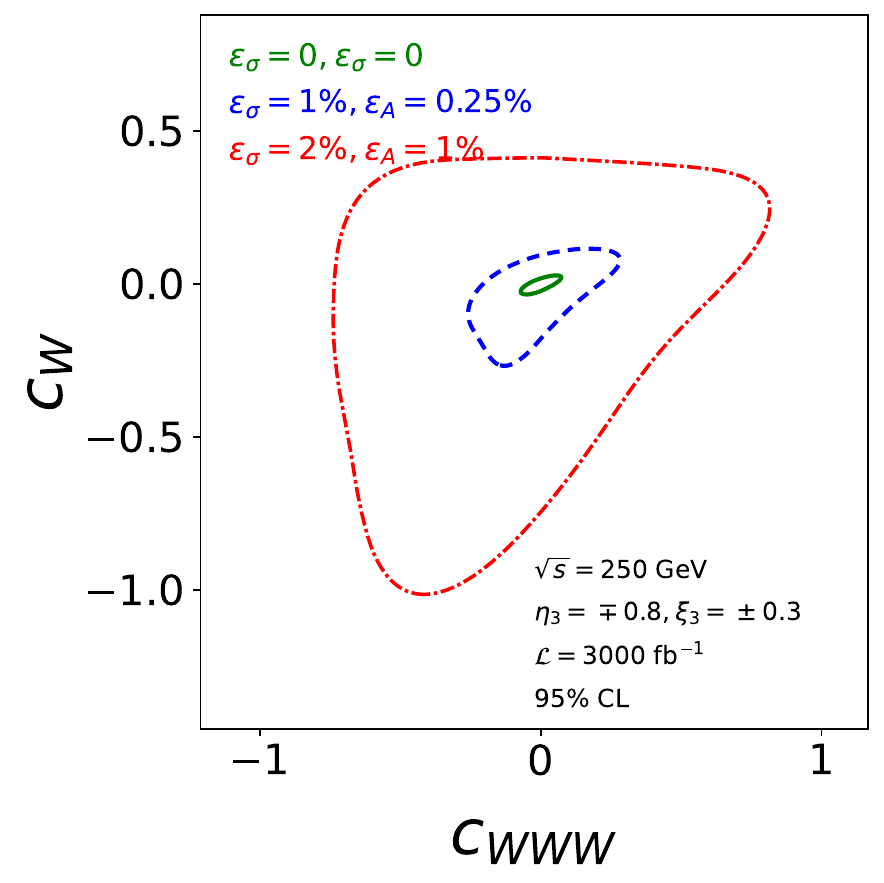}
		\includegraphics[width=0.32\textwidth]{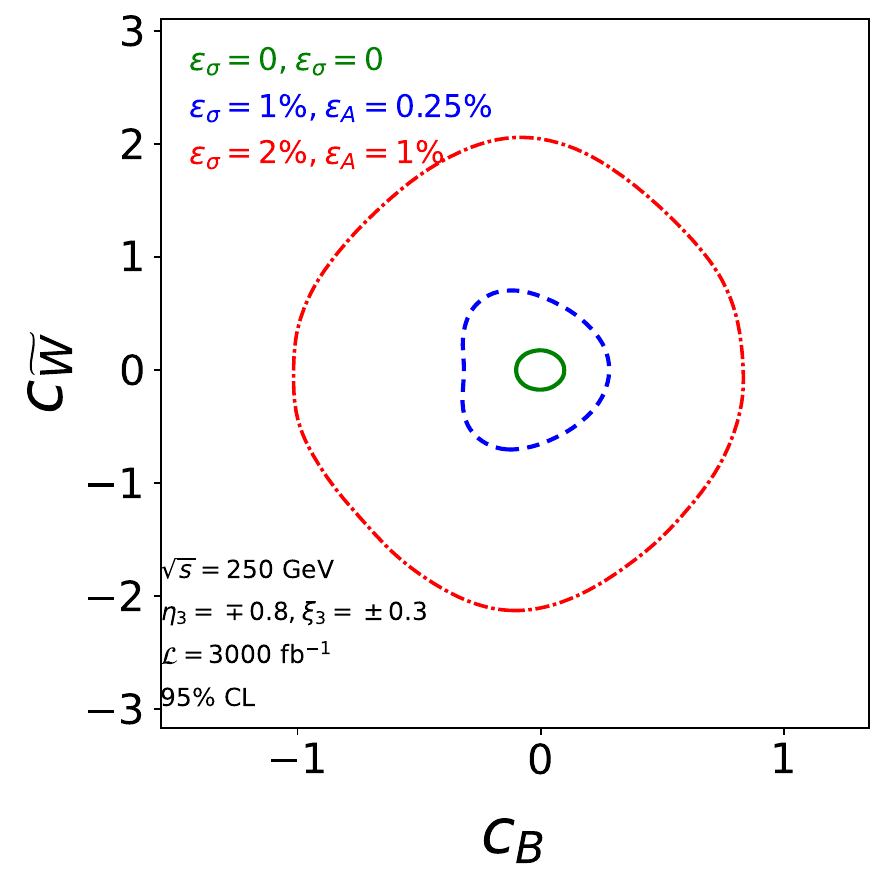}
		\includegraphics[width=0.32\textwidth]{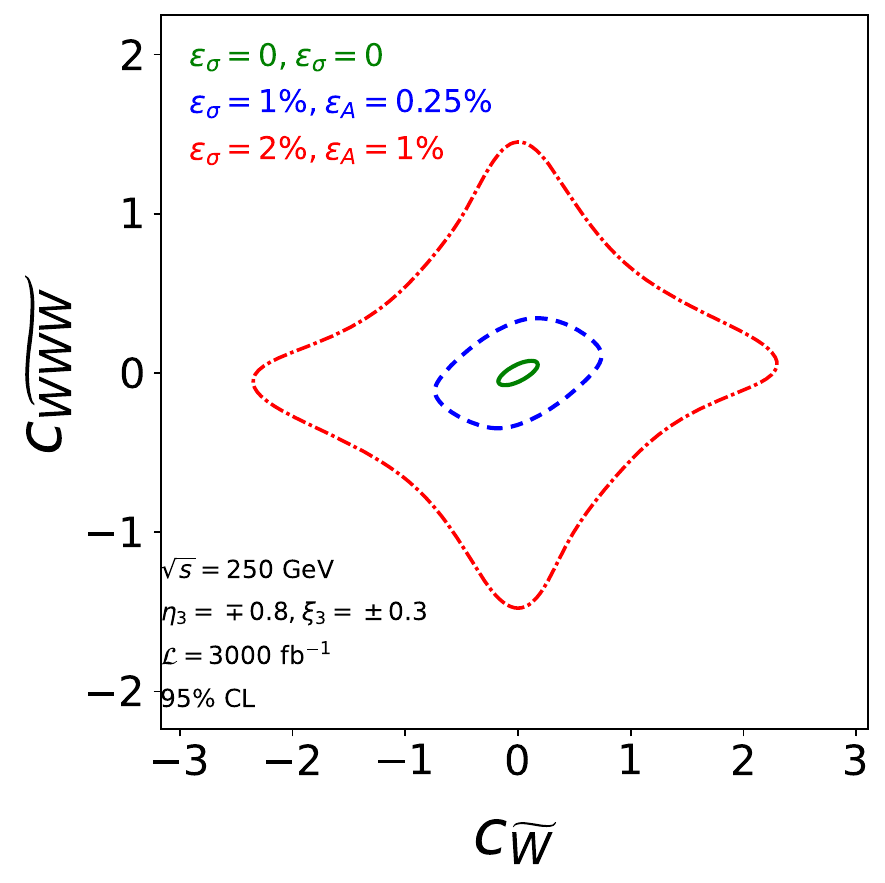}
		\caption{Tow dimensional marginalized projections of the anomalous couplings $c_i$~(TeV$^{-2}$) at $95\%$~CL obtained using the MCMC global fits for a set of systematic errors and integrated luminosity~$\mathcal{L}=3000$~fb$^{-1}$ at $\sqrt{s}=250$~GeV.}
		\label{fig:twod95bci}
	\end{figure*}
	Lastly, we perform a Markov Chain Monte Carlo (MCMC) analysis to obtain a marginalized limits on all five anomalous couplings using binned observables and combining two different sets of initial beam polarizations. To quantify the likelihood of a given point $\mathbf{x} \in \{c_i, \mp\eta_3, \pm\xi_3\}$ in the parameter space, we defined a likelihood function using the chi-squared function,
	\begin{equation}
		\mathcal{L}(\mathbf{x}) \propto \prod_{i,j} \exp\left(-\frac{\chi_{ij}^2(\mathbf{x})}{2}\right),
		\label{eqn:likelihood}
	\end{equation}
	where the indices $i$ and $j$ run over all the bins and observables, respectively. Furthermore, we considered different sets of luminosities, as specified in Eq.~(\ref{eqn:lumi}), and incorporated the systematic errors described in Eq.~(\ref{eqn:syst}). The final luminosity is obtained by doing the analysis at half of the value in Eq.~(\ref{eqn:lumi}) for each set of beam polarization.\\
	The variation of $95\%$ CL marginalized limits of various anomalous couplings~$c_i$ w.r.t luminosities at fixed systematic error is shown in Fig.~\ref{fig:one95}. When the systematic error is zero, increasing the luminosity from $100$ fb$^{-1}$ to $3000$ fb$^{-1}$ leads to tighter limits on all anomalous couplings. This implies that higher luminosities enhance our ability to constrain the values of these couplings, provided systematic are minimized. In the case where systematic errors are present with values of $(\epsilon_\sigma, \epsilon_A) = (1\%, 0.25\%)$, the limits on $c_i$ become tighter as the luminosity increases from $100$~fb$^{-1}$ to $1000$~fb$^{-1}$, however, beyond this point, the limits reach a saturation point, and no significant further improvement is observed. This saturation phenomenon suggests that increasing the luminosity beyond a certain threshold does not yield substantial gains in constraining the couplings when these specific systematic errors are present. On the other hand, for a more conservative estimate of systematic errors with values of $(\epsilon_\sigma, \epsilon_A) = (2\%, 1\%)$, the limits on the anomalous couplings exhibit minimal change with increasing luminosity. The limits practically saturate at a luminosity of $\mathcal{L} = 300$~fb$^{-1}$. This indicates that, in the presence of larger systematic errors, the limits are primarily governed by these uncertainties rather than statistical errors. Consequently, reducing systematic errors becomes crucial for achieving further improvement in constraining the values of the anomalous couplings.
	\\
	In order to investigate the impact of systematic errors on the behavior of limits, we examine two-dimensional marginalization projections for a pair of anomalous couplings. The marginalized projections we consider are pairs of couplings $(c_{WWW},c_W)$, $(c_B,c_{\widetilde{W}})$, and $(c_{\widetilde{W}},c_{\widetilde{WWW}})$, which have been obtained through MCMC global fits at a confidence level of $95\%$ CL. As depicted in Fig.~\ref{fig:twod95bci}, the resulting projections exhibit notable changes when systematic errors are reduced from a conservative level of ($2\%,1\%$). In particular, the contours displayed in the figures undergo significant shrinkage as the systematic errors decrease.\\
	\begin{table*}[!htb]
		\centering
		\renewcommand{\arraystretch}{1.3} 
		\caption{\label{Tab:final} Simultaneous 95$\%$ C.L. limits on the anomalous couplings~$c_i$~(TeV$^{-2}$) obtained with  different values of integrated luminosity in Eq.~(\ref{eqn:lumi}) and systematic error as given in Eq.~(\ref{eqn:syst}). The limits are obtained for $\sqrt{s}=250$~GeV, with two set of beam polarization $(\eta_3,\xi_3) = (\mp0.8,\pm0.3)$.}
		\begin{tabular}{@{}p{2.0cm}@{}p{2.5cm}@{}p{2.7cm}@{}p{2.7cm}@{}p{2.7cm}@{}p{2.7cm}@{}p{2.7cm}}\hline\hline
			$\mathcal{L}$ (fb$^{-1}$)&($\epsilon_\sigma,\epsilon_{\mathcal{A}}$)&\quad$c_{WWW}$&\quad\quad$c_W$&\quad\quad$c_B$&\quad\quad$c_{\widetilde{W}}$&\quad$c_{\widetilde{WWW}}$\\ \hline	
			&$(0.0,0.0)$&$[-0.31,+0.25]$&$[-0.30,+0.10]$&$[-0.53,+0.37]$&$[-0.81,+0.81]$&$[-0.31,+0.31]$\\
			100&$(1\%,0.25\%)$&$[-0.41,+0.31]$&$[-0.33,+0.13]$&$[-0.55,+0.42]$&$[-0.90,+0.90]$&$[-0.41,+0.41]$\\
			&$(2\%,1\%)$&$[-0.75,+0.59]$&$[-0.85,+0.40]$&$[-0.96,+0.79]$&$[-1.74,+1.74]$&$[-1.13,+1.13]$\\
			\hline
			&$(0.0,0.0)$&$[-0.18,+0.17]$&$[-0.13,+0.07]$&$[-0.29,+0.22]$&$[-0.46,+0.46]$&$[-0.19,+0.19]$\\
			300&$(1\%,0.25\%)$&$[-0.30,+0.24]$&$[-0.32,+0.12]$&$[-0.48,+0.31]$&$[-0.80,+0.81]$&$[-0.35,+0.35]$\\
			&$(2\%,1\%)$&$[-0.69,+0.57]$&$[-0.80,+0.39]$&$[-0.86,+0.69]$&$[-1.68,+1.65]$&$[-1.06,+1.06]$\\
			\hline
			&$(0.0,0.0)$&$[-0.10,+0.09]$&$[-0.05,+0.04]$&$[-0.15,+0.13]$&$[-0.24,+0.24]$&$[-0.11,+0.11]$\\
			1000&$(1\%,0.25\%)$&$[-0.24,+0.20]$&$[-0.22,+0.10]$&$[-0.34,+0.23]$&$[-0.62,+0.63]$&$[-0.29,+0.29]$\\
			&$(2\%,1\%)$&$[-0.67,+0.56]$&$[-0.77,+0.38]$&$[-0.82,+0.64]$&$[-1.62,+1.58]$&$[-1.04,+1.02]$\\
			\hline
			&$(0.0,0.0)$&$[-0.06,+0.05]$&$[-0.03,+0.02]$&$[-0.08,+0.07]$&$[-0.13,+0.13]$&$[-0.06,+0.06]$\\
			3000&$(1\%,0.25\%)$&$[-0.22,+0.19]$&$[-0.19,+0.10]$&$[-0.29,+0.20]$&$[-0.56,+0.56]$&$[-0.27,+0.27]$\\
			&$(2\%,1\%)$&$[-0.65,+0.56]$&$[-0.76,+0.38]$&$[-0.81,+0.63]$&$[-1.62,+1.57]$&$[-1.03,+1.02]$\\
			\hline\hline
		\end{tabular}
	\end{table*}
	As a final result, we obtain $95\%$ CL limits on $c_i$ for different luminosities and systematic errors and list the subsequent limits in Table.~\ref{Tab:final}. The limits for $c_{WWW}$ tighten by a factor of $5$ on increasing the $\mathcal{L}$ from $100$ to $3000$~fb$^{-1}$ in case of zero systematic, while in the case of $(1\%,0.25\%)$, and $(2\%,1\%)$ of systematic errors, $c_{WWW}$ tightens by a factor of $\approx 2$, and $1.15$, respectively. Regarding the coupling $c_W$, we find that the lower bounds improve by a factor of $10$, while the upper bounds improve by a factor of $5$ without considering systematic errors. In the presence of systematic errors at levels of $(1\%, 0.2\%)$ and $(2\%, 1\%)$, the lower bounds on $c_W$ tighten by factors of approximately $1.7$ and $1.12$, respectively, while the upper bounds tighten by factors of $1.3$ and $1.1$, respectively. The limits on the couplings $c_B$, $c_{\widetilde{W}}$, and $c_{\widetilde{WWW}}$ exhibit similar behavior. For a conservative estimate of systematic error at $(2\%, 1\%)$, the limits improve by approximately 1.2, 1.1, and 1.1, respectively, when increasing the luminosity from 100~fb$^{-1}$ to 3000~fb$^{-1}$ The tightening of limits of all the anomalous couplings saturates to a factor of approximately 1.1 for systematic error of $(2\%,1\%)$ even when the luminosity increases by a factor of 30. Thus, reducing systematic errors becomes paramount in probing new physics.
	\\\\
	At this juncture, we want to compare the marginalized limits obtained in full-hadronic with the semi-leptonic channel~\cite{Subba:2023rpm}. For an integrated luminosity of $100$~fb$^{-1}$ and systematic error $(2\%,1\%)$, the current~(full-hadronic) lower limits on $c_{WWW}$ is tighter by a factor of $\approx$ $2.41$, while upper bounds become tighter by a factor of $3.0$. Similarly, the limits on $c_{\widetilde{W}}$ and $c_{\widetilde{WWW}}$ in full-hadronic channel get tighter by a factor of $3.33$ and $1.42$, respectively in comparison to semi-leptonic limits. The lower bound of $c_W$, and $c_B$ gets tighter by a factor of $1.53$, and $1.67$, while the upper bound shrink by a factor of $3.25$, and $2.28$, respectively. Thus there is a significant improvement in the credible limits of anomalous couplings in comparison to semi-leptonic channel. The huge improvement in the limits on anomalous couplings results due to large interference between the {\tt Non-WW} zero-resonant and {\tt WW} amplitudes of {\tt CH2}.
	
	\section{Conclusion}
	\label{sec:conclude}
	The full-hadronic decay of the $W$ boson in the $W^-W^+$ di-boson production process is studied at leading order at $\sqrt{s}=250$~GeV with initial polarized beams. This hadronic channel becomes important owing to the large cross~section and significant negative interference between the double resonant ${\tt WW}$ and zero resonant ${\tt Non-WW}$ topology. We perform the reconstruction of two $W$ boson using the jet charge variable, and the flavor identity of final jets of tagged $W$ boson as \emph{up/down}-type are achieved using boosted decision trees. The flavor identity of final jets were needed to reconstruct vector polarization and its related spin-correlation asymmetries. The spin-related observables along with cross~section are used to constrain the anomalous couplings affecting $W^-W^+\gamma/Z$ vertex.
	
	The one parameter limits obtained in this work for all five anomalous couplings are tighter than the corresponding experimental limits, as is listed in Table~\ref{tab:Tab1}. The corresponding bounds on $c_{WWW}$,and $c_B$  are tighter by a factor of $3.3$, and $25.8$, respectively. For $c_W$, the upper and lower bounds were improved by a factor of $4.3$, and $30.0$ in comparison to experimental limits. Similarly, the limits on $CP$-odd couplings $c_{\widetilde{W}},$ and $c_{\widetilde{WWW}}$ get tighter by a factor of $28.6$, and $1.4$, respectively. A notable fact is that in the case of $c_{\widetilde{WWW}}$, the sensitivity of cross~section to this couplings directly increases with center of mass energy due to momentum term in the anomalous vertex. This improved sensitivity is exploited at LHC running with high energy~$(13~\text{TeV})$, yet the current results showed that even for the machine with much lower energy, $\sqrt{s}=250$~GeV, the full-hadronic channel offers significant constraint on those parameters. It is due to the large number of sensitive spin-related observables. 
	
	The increase in sensitivity due to the interference between doubly resonant ${\tt WW}$ with the zero resonant {\tt Non-WW} amplitudes of {\tt CH2} are translated to the bounds on anomalous couplings. It is clearly highlighted in Table~\ref{tab:Tab1}, where the comparison can be made with the semi-leptonic channel where such interference effects are absent. We note a significant improvement on all five anomalous couplings in full-hadronic as compare to semi-leptonic one. In comparison to the semi-leptonic channel, the limits on $c_i\in \{c_{WWW},c_W,c_B,c_{\widetilde{W}},c_{\widetilde{W}WW}$\} set from full-hadronic channel are tighter by a factor ranging from $3$ to $9$.
	\begin{acknowledgments}
		A.~Subba acknowledge the financial support of the University Grant Commission, Govt. of India through UGC-NET Senior Research Fellowship.
	\end{acknowledgments}

	\appendix
	\section{Decay density matrix of $W$ boson}
	\label{sec:sdm}
	For the decay of spin-1 $W$ boson decaying to $f\bar{f}$ pairs with decay vertex $\bar{f}\gamma^\mu P_L f V$, the decay density matrix is given by~\cite{Boudjema:2009fz},
	\begin{widetext}
		\begin{equation}
			\Gamma_W(\lambda_W,\lambda_W^\prime) =
			\begin{bmatrix}
				\frac{1+\delta+(1-3\delta)\cos^2\theta+2\alpha\cos\theta}{4}&\frac{\sin\theta(\alpha+(1-3\delta)\cos\theta)}{2\sqrt{2}}e^{i\phi}&(1-3\delta)\frac{(1-\cos^2\theta)}{4}e^{i2\phi}\\
				\frac{\sin\theta(\alpha+(1-3\delta)\cos\theta)}{2\sqrt{2}}e^{-i\phi}&\delta+(1-3\delta)\frac{\sin^2\theta}{2}&\frac{\sin\theta(\alpha-(1-3\delta)\cos\theta)}{2\sqrt{s}}e^{i\phi}\\
				(1-3\delta)\frac{(1-\cos^2\theta)}{4}e^{-i2\phi}&\frac{\sin\theta(\alpha-(1-3\delta)\cos\theta)}{2\sqrt{2}}e^{-i\phi}&\frac{1+\delta+(1-3\delta)\cos^2\theta-2\alpha\cos\theta}{4}
			\end{bmatrix},
		\end{equation}
	\end{widetext}
	where the $\theta$, and $\phi$ are the polar and azimuth orientation of final decayed fermions at the rest frame of $W$ boson.

	Here the spin analyzing power $\alpha$ is given by,
		\begin{equation*}
		\frac{2(C_R^2-C_L^2)\sqrt{1+(x_1^2-x_2^2)^2-2(x_1^2+x_2^2)}}{12C_LC_Rx_1x_2+(C_R^2+C_L^2)\left[2-(x_1^2-x_2^2)^2+(x_1^2+x_2^2)\right]}.
	\end{equation*}
	The parameter $\delta$ for the case $W$ boson decaying to two fermions is defined as,
	\begin{equation*}
		\frac{4C_LC_Rx_1x_2+(C_R^2+C_L^2)\left[(x_1^2+x_2^2)-(x_1^2-x_2^2)^2\right]}{12C_LC_Rx_!x_2+(C_R^2+C_L^2)\left[2-(x_1^2-x_2^2)^2+(x_1^2+x_2^2)\right]},
	\end{equation*}
	where $x_i = m_i/M$ with $m_i$ the mass of final jets and $M$ the mass of $W$ boson. At the high energy limit, $x_i \to 0$, and $\alpha \to (C_R^2-C_L^2)/(C_R^2+C_L^2)$, and $\delta \to 0$. Within SM at the leading order, we have $C_R = 0$ hence $\alpha = -1$. 
	\section{List of features used for flavor tagging}
	\label{sec:app2}
	For completeness, we list a set of features that were used as an input to BDT networks for flavor tagging. The features are,
	\begin{itemize}[itemsep=0.8pt]
		\item Leptons: Counts, 4-momenta,
		\item Total number of visible particles,
		\item Count of positive and negative charged particles,
		\item Charged Kaons: Counts, 4-momenta,
		\item Charged Pions: Counts, 4-momenta,
		\item Total number of positive and negative charged hadrons,
		\item  Displaced tracks satisfying $p_T > 1.0$~GeV are selected. They are binned with respect to the lifetime~($\tau$) in $mm$ of their mother particles,
		\begin{itemize}
			\item $D1: 0.3 \le \tau \le 3.0$,
			\item $D2: 3.0 \le \tau \le 30.0$,
			\item $D3: 30.0 \le \tau \le 300.0$,
			\item $D4: 300.0 \le \tau \le 1200.0$,
			\item $D5: \tau \ge 1200.0$
		\end{itemize}
		\item Energy of photons,
		\item Energy of charged hadrons.
	\end{itemize}
	\nocite{*}
	
	\bibliography{refer}

\begin{thebibliography}{61}%
\makeatletter
\providecommand \@ifxundefined [1]{%
 \@ifx{#1\undefined}
}%
\providecommand \@ifnum [1]{%
 \ifnum #1\expandafter \@firstoftwo
 \else \expandafter \@secondoftwo
 \fi
}%
\providecommand \@ifx [1]{%
 \ifx #1\expandafter \@firstoftwo
 \else \expandafter \@secondoftwo
 \fi
}%
\providecommand \natexlab [1]{#1}%
\providecommand \enquote  [1]{``#1''}%
\providecommand \bibnamefont  [1]{#1}%
\providecommand \bibfnamefont [1]{#1}%
\providecommand \citenamefont [1]{#1}%
\providecommand \href@noop [0]{\@secondoftwo}%
\providecommand \href [0]{\begingroup \@sanitize@url \@href}%
\providecommand \@href[1]{\@@startlink{#1}\@@href}%
\providecommand \@@href[1]{\endgroup#1\@@endlink}%
\providecommand \@sanitize@url [0]{\catcode `\\12\catcode `\$12\catcode
  `\&12\catcode `\#12\catcode `\^12\catcode `\_12\catcode `\%12\relax}%
\providecommand \@@startlink[1]{}%
\providecommand \@@endlink[0]{}%
\providecommand \url  [0]{\begingroup\@sanitize@url \@url }%
\providecommand \@url [1]{\endgroup\@href {#1}{\urlprefix }}%
\providecommand \urlprefix  [0]{URL }%
\providecommand \Eprint [0]{\href }%
\providecommand \doibase [0]{https://doi.org/}%
\providecommand \selectlanguage [0]{\@gobble}%
\providecommand \bibinfo  [0]{\@secondoftwo}%
\providecommand \bibfield  [0]{\@secondoftwo}%
\providecommand \translation [1]{[#1]}%
\providecommand \BibitemOpen [0]{}%
\providecommand \bibitemStop [0]{}%
\providecommand \bibitemNoStop [0]{.\EOS\space}%
\providecommand \EOS [0]{\spacefactor3000\relax}%
\providecommand \BibitemShut  [1]{\csname bibitem#1\endcsname}%
\let\auto@bib@innerbib\@empty
\bibitem [{\citenamefont {Chatrchyan}\ \emph {et~al.}(2012)\citenamefont
  {Chatrchyan} \emph {et~al.}}]{CMS:2012qbp}%
  \BibitemOpen
  \bibfield  {author} {\bibinfo {author} {\bibfnamefont {S.}~\bibnamefont
  {Chatrchyan}} \emph {et~al.} (\bibinfo {collaboration} {CMS}),\ }\bibfield
  {title} {\bibinfo {title} {{Observation of a New Boson at a Mass of 125 GeV
  with the CMS Experiment at the LHC}},\ }\href
  {https://doi.org/10.1016/j.physletb.2012.08.021} {\bibfield  {journal}
  {\bibinfo  {journal} {Phys. Lett. B}\ }\textbf {\bibinfo {volume} {716}},\
  \bibinfo {pages} {30} (\bibinfo {year} {2012})},\ \Eprint
  {https://arxiv.org/abs/1207.7235} {arXiv:1207.7235 [hep-ex]} \BibitemShut
  {NoStop}%
\bibitem [{\citenamefont {Aad}\ \emph {et~al.}(2012{\natexlab{a}})\citenamefont
  {Aad} \emph {et~al.}}]{ATLAS:2012yve}%
  \BibitemOpen
  \bibfield  {author} {\bibinfo {author} {\bibfnamefont {G.}~\bibnamefont
  {Aad}} \emph {et~al.} (\bibinfo {collaboration} {ATLAS}),\ }\bibfield
  {title} {\bibinfo {title} {{Observation of a new particle in the search for
  the Standard Model Higgs boson with the ATLAS detector at the LHC}},\ }\href
  {https://doi.org/10.1016/j.physletb.2012.08.020} {\bibfield  {journal}
  {\bibinfo  {journal} {Phys. Lett. B}\ }\textbf {\bibinfo {volume} {716}},\
  \bibinfo {pages} {1} (\bibinfo {year} {2012}{\natexlab{a}})},\ \Eprint
  {https://arxiv.org/abs/1207.7214} {arXiv:1207.7214 [hep-ex]} \BibitemShut
  {NoStop}%
\bibitem [{\citenamefont {Higgs}(1964)}]{Higgs:1964pj}%
  \BibitemOpen
  \bibfield  {author} {\bibinfo {author} {\bibfnamefont {P.~W.}\ \bibnamefont
  {Higgs}},\ }\bibfield  {title} {\bibinfo {title} {{Broken Symmetries and the
  Masses of Gauge Bosons}},\ }\href
  {https://doi.org/10.1103/PhysRevLett.13.508} {\bibfield  {journal} {\bibinfo
  {journal} {Phys. Rev. Lett.}\ }\textbf {\bibinfo {volume} {13}},\ \bibinfo
  {pages} {508} (\bibinfo {year} {1964})}\BibitemShut {NoStop}%
\bibitem [{\citenamefont {Englert}\ and\ \citenamefont
  {Brout}(1964)}]{Englert:1964et}%
  \BibitemOpen
  \bibfield  {author} {\bibinfo {author} {\bibfnamefont {F.}~\bibnamefont
  {Englert}}\ and\ \bibinfo {author} {\bibfnamefont {R.}~\bibnamefont
  {Brout}},\ }\bibfield  {title} {\bibinfo {title} {{Broken Symmetry and the
  Mass of Gauge Vector Mesons}},\ }\href
  {https://doi.org/10.1103/PhysRevLett.13.321} {\bibfield  {journal} {\bibinfo
  {journal} {Phys. Rev. Lett.}\ }\textbf {\bibinfo {volume} {13}},\ \bibinfo
  {pages} {321} (\bibinfo {year} {1964})}\BibitemShut {NoStop}%
\bibitem [{\citenamefont {Aad}\ \emph {et~al.}(2021)\citenamefont {Aad} \emph
  {et~al.}}]{ATLAS:2021jgw}%
  \BibitemOpen
  \bibfield  {author} {\bibinfo {author} {\bibfnamefont {G.}~\bibnamefont
  {Aad}} \emph {et~al.} (\bibinfo {collaboration} {ATLAS}),\ }\bibfield
  {title} {\bibinfo {title} {{Measurements of $W^+W^-+\ge 1~$jet production
  cross-sections in $pp$ collisions at $\sqrt{s}=13~$TeV with the ATLAS
  detector}},\ }\href {https://doi.org/10.1007/JHEP06(2021)003} {\bibfield
  {journal} {\bibinfo  {journal} {JHEP}\ }\textbf {\bibinfo {volume} {06}},\
  \bibinfo {pages} {003}},\ \Eprint {https://arxiv.org/abs/2103.10319}
  {arXiv:2103.10319 [hep-ex]} \BibitemShut {NoStop}%
\bibitem [{\citenamefont {Chatrchyan}\ \emph {et~al.}(2013)\citenamefont
  {Chatrchyan} \emph {et~al.}}]{CMS:2013ant}%
  \BibitemOpen
  \bibfield  {author} {\bibinfo {author} {\bibfnamefont {S.}~\bibnamefont
  {Chatrchyan}} \emph {et~al.} (\bibinfo {collaboration} {CMS}),\ }\bibfield
  {title} {\bibinfo {title} {{Measurement of the $W^+W^-$ Cross Section in $pp$
  Collisions at $\sqrt{s} = 7$ TeV and Limits on Anomalous $WW\gamma$ and $WWZ$
  Couplings}},\ }\href {https://doi.org/10.1140/epjc/s10052-013-2610-8}
  {\bibfield  {journal} {\bibinfo  {journal} {Eur. Phys. J. C}\ }\textbf
  {\bibinfo {volume} {73}},\ \bibinfo {pages} {2610} (\bibinfo {year}
  {2013})},\ \Eprint {https://arxiv.org/abs/1306.1126} {arXiv:1306.1126
  [hep-ex]} \BibitemShut {NoStop}%
\bibitem [{\citenamefont {Aad}\ \emph {et~al.}(2012{\natexlab{b}})\citenamefont
  {Aad} \emph {et~al.}}]{ATLAS:2012upi}%
  \BibitemOpen
  \bibfield  {author} {\bibinfo {author} {\bibfnamefont {G.}~\bibnamefont
  {Aad}} \emph {et~al.} (\bibinfo {collaboration} {ATLAS}),\ }\bibfield
  {title} {\bibinfo {title} {{Measurement of the $W W$ cross section in
  $\sqrt{s}=7$ TeV $pp$ collisions with the ATLAS detector and limits on
  anomalous gauge couplings}},\ }\href
  {https://doi.org/10.1016/j.physletb.2012.05.003} {\bibfield  {journal}
  {\bibinfo  {journal} {Phys. Lett. B}\ }\textbf {\bibinfo {volume} {712}},\
  \bibinfo {pages} {289} (\bibinfo {year} {2012}{\natexlab{b}})},\ \Eprint
  {https://arxiv.org/abs/1203.6232} {arXiv:1203.6232 [hep-ex]} \BibitemShut
  {NoStop}%
\bibitem [{\citenamefont {Tumasyan}\ \emph {et~al.}(2022)\citenamefont
  {Tumasyan} \emph {et~al.}}]{CMS:2021icx}%
  \BibitemOpen
  \bibfield  {author} {\bibinfo {author} {\bibfnamefont {A.}~\bibnamefont
  {Tumasyan}} \emph {et~al.} (\bibinfo {collaboration} {CMS}),\ }\bibfield
  {title} {\bibinfo {title} {{Measurement of the inclusive and differential WZ
  production cross sections, polarization angles, and triple gauge couplings in
  pp collisions at $ \sqrt{s} $ = 13 TeV}},\ }\href
  {https://doi.org/10.1007/JHEP07(2022)032} {\bibfield  {journal} {\bibinfo
  {journal} {JHEP}\ }\textbf {\bibinfo {volume} {07}},\ \bibinfo {pages}
  {032}},\ \Eprint {https://arxiv.org/abs/2110.11231} {arXiv:2110.11231
  [hep-ex]} \BibitemShut {NoStop}%
\bibitem [{\citenamefont {Sirunyan}\ \emph
  {et~al.}(2019{\natexlab{a}})\citenamefont {Sirunyan} \emph
  {et~al.}}]{CMS:2019efc}%
  \BibitemOpen
  \bibfield  {author} {\bibinfo {author} {\bibfnamefont {A.~M.}\ \bibnamefont
  {Sirunyan}} \emph {et~al.} (\bibinfo {collaboration} {CMS}),\ }\bibfield
  {title} {\bibinfo {title} {{Measurements of the pp $\to$ WZ inclusive and
  differential production cross section and constraints on charged anomalous
  triple gauge couplings at $\sqrt{s} =$ 13 TeV}},\ }\href
  {https://doi.org/10.1007/JHEP04(2019)122} {\bibfield  {journal} {\bibinfo
  {journal} {JHEP}\ }\textbf {\bibinfo {volume} {04}},\ \bibinfo {pages}
  {122}},\ \Eprint {https://arxiv.org/abs/1901.03428} {arXiv:1901.03428
  [hep-ex]} \BibitemShut {NoStop}%
\bibitem [{\citenamefont {Aad}\ \emph {et~al.}(2016{\natexlab{a}})\citenamefont
  {Aad} \emph {et~al.}}]{ATLAS:2016qjc}%
  \BibitemOpen
  \bibfield  {author} {\bibinfo {author} {\bibfnamefont {G.}~\bibnamefont
  {Aad}} \emph {et~al.} (\bibinfo {collaboration} {ATLAS}),\ }\bibfield
  {title} {\bibinfo {title} {{Measurements of $Z\gamma$ and $Z\gamma\gamma$
  production in $pp$ collisions at $\sqrt{s}=$ 8 TeV with the ATLAS
  detector}},\ }\href {https://doi.org/10.1103/PhysRevD.93.112002} {\bibfield
  {journal} {\bibinfo  {journal} {Phys. Rev. D}\ }\textbf {\bibinfo {volume}
  {93}},\ \bibinfo {pages} {112002} (\bibinfo {year} {2016}{\natexlab{a}})},\
  \Eprint {https://arxiv.org/abs/1604.05232} {arXiv:1604.05232 [hep-ex]}
  \BibitemShut {NoStop}%
\bibitem [{\citenamefont {Aad}\ \emph {et~al.}(2016{\natexlab{b}})\citenamefont
  {Aad} \emph {et~al.}}]{ATLAS:2016bkj}%
  \BibitemOpen
  \bibfield  {author} {\bibinfo {author} {\bibfnamefont {G.}~\bibnamefont
  {Aad}} \emph {et~al.} (\bibinfo {collaboration} {ATLAS}),\ }\bibfield
  {title} {\bibinfo {title} {{Measurements of $W^\pm Z$ production cross
  sections in $pp$ collisions at $\sqrt{s} = 8$ TeV with the ATLAS detector and
  limits on anomalous gauge boson self-couplings}},\ }\href
  {https://doi.org/10.1103/PhysRevD.93.092004} {\bibfield  {journal} {\bibinfo
  {journal} {Phys. Rev. D}\ }\textbf {\bibinfo {volume} {93}},\ \bibinfo
  {pages} {092004} (\bibinfo {year} {2016}{\natexlab{b}})},\ \Eprint
  {https://arxiv.org/abs/1603.02151} {arXiv:1603.02151 [hep-ex]} \BibitemShut
  {NoStop}%
\bibitem [{\citenamefont {Chatrchyan}\ \emph {et~al.}(2014)\citenamefont
  {Chatrchyan} \emph {et~al.}}]{CMS:2013ryd}%
  \BibitemOpen
  \bibfield  {author} {\bibinfo {author} {\bibfnamefont {S.}~\bibnamefont
  {Chatrchyan}} \emph {et~al.} (\bibinfo {collaboration} {CMS}),\ }\bibfield
  {title} {\bibinfo {title} {{Measurement of the $W\gamma$ and $Z\gamma$
  Inclusive Cross Sections in $pp$ Collisions at $\sqrt s=7$ TeV and Limits on
  Anomalous Triple Gauge Boson Couplings}},\ }\href
  {https://doi.org/10.1103/PhysRevD.89.092005} {\bibfield  {journal} {\bibinfo
  {journal} {Phys. Rev. D}\ }\textbf {\bibinfo {volume} {89}},\ \bibinfo
  {pages} {092005} (\bibinfo {year} {2014})},\ \Eprint
  {https://arxiv.org/abs/1308.6832} {arXiv:1308.6832 [hep-ex]} \BibitemShut
  {NoStop}%
\bibitem [{\citenamefont {Aad}\ \emph {et~al.}(2013)\citenamefont {Aad} \emph
  {et~al.}}]{ATLAS:2013way}%
  \BibitemOpen
  \bibfield  {author} {\bibinfo {author} {\bibfnamefont {G.}~\bibnamefont
  {Aad}} \emph {et~al.} (\bibinfo {collaboration} {ATLAS}),\ }\bibfield
  {title} {\bibinfo {title} {{Measurements of $W \gamma$ and $Z \gamma$
  production in $pp$ collisions at $\sqrt{s}$=7 TeV with the ATLAS detector at
  the LHC}},\ }\href {https://doi.org/10.1103/PhysRevD.87.112003} {\bibfield
  {journal} {\bibinfo  {journal} {Phys. Rev. D}\ }\textbf {\bibinfo {volume}
  {87}},\ \bibinfo {pages} {112003} (\bibinfo {year} {2013})},\ \bibinfo {note}
  {[Erratum: Phys.Rev.D 91, 119901 (2015)]},\ \Eprint
  {https://arxiv.org/abs/1302.1283} {arXiv:1302.1283 [hep-ex]} \BibitemShut
  {NoStop}%
\bibitem [{\citenamefont {Aad}\ \emph {et~al.}(2012{\natexlab{c}})\citenamefont
  {Aad} \emph {et~al.}}]{ATLAS:2012bpb}%
  \BibitemOpen
  \bibfield  {author} {\bibinfo {author} {\bibfnamefont {G.}~\bibnamefont
  {Aad}} \emph {et~al.} (\bibinfo {collaboration} {ATLAS}),\ }\bibfield
  {title} {\bibinfo {title} {{Measurement of $W \gamma$ and $Z \gamma$
  production cross sections in $pp$ collisions at $\sqrt{s}=7$ TeV and limits
  on anomalous triple gauge couplings with the ATLAS detector}},\ }\href
  {https://doi.org/10.1016/j.physletb.2012.09.017} {\bibfield  {journal}
  {\bibinfo  {journal} {Phys. Lett. B}\ }\textbf {\bibinfo {volume} {717}},\
  \bibinfo {pages} {49} (\bibinfo {year} {2012}{\natexlab{c}})},\ \Eprint
  {https://arxiv.org/abs/1205.2531} {arXiv:1205.2531 [hep-ex]} \BibitemShut
  {NoStop}%
\bibitem [{\citenamefont {Aaltonen}\ \emph {et~al.}(2012)\citenamefont
  {Aaltonen} \emph {et~al.}}]{CDF:2012mnr}%
  \BibitemOpen
  \bibfield  {author} {\bibinfo {author} {\bibfnamefont {T.}~\bibnamefont
  {Aaltonen}} \emph {et~al.} (\bibinfo {collaboration} {CDF}),\ }\bibfield
  {title} {\bibinfo {title} {{Measurement of the $WZ$ Cross Section and Triple
  Gauge Couplings in $p \bar p$ Collisions at $\sqrt{s} = 1.96$ TeV}},\ }\href
  {https://doi.org/10.1103/PhysRevD.86.031104} {\bibfield  {journal} {\bibinfo
  {journal} {Phys. Rev. D}\ }\textbf {\bibinfo {volume} {86}},\ \bibinfo
  {pages} {031104} (\bibinfo {year} {2012})},\ \Eprint
  {https://arxiv.org/abs/1202.6629} {arXiv:1202.6629 [hep-ex]} \BibitemShut
  {NoStop}%
\bibitem [{\citenamefont {Hagiwara}\ \emph {et~al.}(1987)\citenamefont
  {Hagiwara}, \citenamefont {Peccei}, \citenamefont {Zeppenfeld},\ and\
  \citenamefont {Hikasa}}]{Hagiwara:1986vm}%
  \BibitemOpen
  \bibfield  {author} {\bibinfo {author} {\bibfnamefont {K.}~\bibnamefont
  {Hagiwara}}, \bibinfo {author} {\bibfnamefont {R.~D.}\ \bibnamefont
  {Peccei}}, \bibinfo {author} {\bibfnamefont {D.}~\bibnamefont {Zeppenfeld}},\
  and\ \bibinfo {author} {\bibfnamefont {K.}~\bibnamefont {Hikasa}},\
  }\bibfield  {title} {\bibinfo {title} {{Probing the Weak Boson Sector in e+
  e- ---\ensuremath{>} W+ W-}},\ }\href
  {https://doi.org/10.1016/0550-3213(87)90685-7} {\bibfield  {journal}
  {\bibinfo  {journal} {Nucl. Phys. B}\ }\textbf {\bibinfo {volume} {282}},\
  \bibinfo {pages} {253} (\bibinfo {year} {1987})}\BibitemShut {NoStop}%
\bibitem [{\citenamefont {Degrande}\ \emph {et~al.}(2013)\citenamefont
  {Degrande}, \citenamefont {Greiner}, \citenamefont {Kilian}, \citenamefont
  {Mattelaer}, \citenamefont {Mebane}, \citenamefont {Stelzer}, \citenamefont
  {Willenbrock},\ and\ \citenamefont {Zhang}}]{Degrande:2012wf}%
  \BibitemOpen
  \bibfield  {author} {\bibinfo {author} {\bibfnamefont {C.}~\bibnamefont
  {Degrande}}, \bibinfo {author} {\bibfnamefont {N.}~\bibnamefont {Greiner}},
  \bibinfo {author} {\bibfnamefont {W.}~\bibnamefont {Kilian}}, \bibinfo
  {author} {\bibfnamefont {O.}~\bibnamefont {Mattelaer}}, \bibinfo {author}
  {\bibfnamefont {H.}~\bibnamefont {Mebane}}, \bibinfo {author} {\bibfnamefont
  {T.}~\bibnamefont {Stelzer}}, \bibinfo {author} {\bibfnamefont
  {S.}~\bibnamefont {Willenbrock}},\ and\ \bibinfo {author} {\bibfnamefont
  {C.}~\bibnamefont {Zhang}},\ }\bibfield  {title} {\bibinfo {title}
  {{Effective Field Theory: A Modern Approach to Anomalous Couplings}},\ }\href
  {https://doi.org/10.1016/j.aop.2013.04.016} {\bibfield  {journal} {\bibinfo
  {journal} {Annals Phys.}\ }\textbf {\bibinfo {volume} {335}},\ \bibinfo
  {pages} {21} (\bibinfo {year} {2013})},\ \Eprint
  {https://arxiv.org/abs/1205.4231} {arXiv:1205.4231 [hep-ph]} \BibitemShut
  {NoStop}%
\bibitem [{\citenamefont {Buchmuller}\ and\ \citenamefont
  {Wyler}(1986)}]{Buchmuller:1985jz}%
  \BibitemOpen
  \bibfield  {author} {\bibinfo {author} {\bibfnamefont {W.}~\bibnamefont
  {Buchmuller}}\ and\ \bibinfo {author} {\bibfnamefont {D.}~\bibnamefont
  {Wyler}},\ }\bibfield  {title} {\bibinfo {title} {{Effective Lagrangian
  Analysis of New Interactions and Flavor Conservation}},\ }\href
  {https://doi.org/10.1016/0550-3213(86)90262-2} {\bibfield  {journal}
  {\bibinfo  {journal} {Nucl. Phys. B}\ }\textbf {\bibinfo {volume} {268}},\
  \bibinfo {pages} {621} (\bibinfo {year} {1986})}\BibitemShut {NoStop}%
\bibitem [{\citenamefont {Grzadkowski}\ \emph {et~al.}(2010)\citenamefont
  {Grzadkowski}, \citenamefont {Iskrzynski}, \citenamefont {Misiak},\ and\
  \citenamefont {Rosiek}}]{Grzadkowski:2010es}%
  \BibitemOpen
  \bibfield  {author} {\bibinfo {author} {\bibfnamefont {B.}~\bibnamefont
  {Grzadkowski}}, \bibinfo {author} {\bibfnamefont {M.}~\bibnamefont
  {Iskrzynski}}, \bibinfo {author} {\bibfnamefont {M.}~\bibnamefont {Misiak}},\
  and\ \bibinfo {author} {\bibfnamefont {J.}~\bibnamefont {Rosiek}},\
  }\bibfield  {title} {\bibinfo {title} {{Dimension-Six Terms in the Standard
  Model Lagrangian}},\ }\href {https://doi.org/10.1007/JHEP10(2010)085}
  {\bibfield  {journal} {\bibinfo  {journal} {JHEP}\ }\textbf {\bibinfo
  {volume} {10}},\ \bibinfo {pages} {085}},\ \Eprint
  {https://arxiv.org/abs/1008.4884} {arXiv:1008.4884 [hep-ph]} \BibitemShut
  {NoStop}%
\bibitem [{\citenamefont {Brivio}\ and\ \citenamefont
  {Trott}(2019)}]{Brivio:2017vri}%
  \BibitemOpen
  \bibfield  {author} {\bibinfo {author} {\bibfnamefont {I.}~\bibnamefont
  {Brivio}}\ and\ \bibinfo {author} {\bibfnamefont {M.}~\bibnamefont {Trott}},\
  }\bibfield  {title} {\bibinfo {title} {{The Standard Model as an Effective
  Field Theory}},\ }\href {https://doi.org/10.1016/j.physrep.2018.11.002}
  {\bibfield  {journal} {\bibinfo  {journal} {Phys. Rept.}\ }\textbf {\bibinfo
  {volume} {793}},\ \bibinfo {pages} {1} (\bibinfo {year} {2019})},\ \Eprint
  {https://arxiv.org/abs/1706.08945} {arXiv:1706.08945 [hep-ph]} \BibitemShut
  {NoStop}%
\bibitem [{\citenamefont {Hagiwara}\ \emph {et~al.}(1993)\citenamefont
  {Hagiwara}, \citenamefont {Ishihara}, \citenamefont {Szalapski},\ and\
  \citenamefont {Zeppenfeld}}]{Hagiwara:1993ck}%
  \BibitemOpen
  \bibfield  {author} {\bibinfo {author} {\bibfnamefont {K.}~\bibnamefont
  {Hagiwara}}, \bibinfo {author} {\bibfnamefont {S.}~\bibnamefont {Ishihara}},
  \bibinfo {author} {\bibfnamefont {R.}~\bibnamefont {Szalapski}},\ and\
  \bibinfo {author} {\bibfnamefont {D.}~\bibnamefont {Zeppenfeld}},\ }\bibfield
   {title} {\bibinfo {title} {{Low-energy effects of new interactions in the
  electroweak boson sector}},\ }\href
  {https://doi.org/10.1103/PhysRevD.48.2182} {\bibfield  {journal} {\bibinfo
  {journal} {Phys. Rev. D}\ }\textbf {\bibinfo {volume} {48}},\ \bibinfo
  {pages} {2182} (\bibinfo {year} {1993})}\BibitemShut {NoStop}%
\bibitem [{\citenamefont {Hagiwara}\ \emph {et~al.}(1992)\citenamefont
  {Hagiwara}, \citenamefont {Ishihara}, \citenamefont {Szalapski},\ and\
  \citenamefont {Zeppenfeld}}]{Hagiwara:1992eh}%
  \BibitemOpen
  \bibfield  {author} {\bibinfo {author} {\bibfnamefont {K.}~\bibnamefont
  {Hagiwara}}, \bibinfo {author} {\bibfnamefont {S.}~\bibnamefont {Ishihara}},
  \bibinfo {author} {\bibfnamefont {R.}~\bibnamefont {Szalapski}},\ and\
  \bibinfo {author} {\bibfnamefont {D.}~\bibnamefont {Zeppenfeld}},\ }\bibfield
   {title} {\bibinfo {title} {{Low-energy constraints on electroweak three
  gauge boson couplings}},\ }\href
  {https://doi.org/10.1016/0370-2693(92)90031-X} {\bibfield  {journal}
  {\bibinfo  {journal} {Phys. Lett. B}\ }\textbf {\bibinfo {volume} {283}},\
  \bibinfo {pages} {353} (\bibinfo {year} {1992})}\BibitemShut {NoStop}%
\bibitem [{\citenamefont {Buchalla}\ \emph {et~al.}(2013)\citenamefont
  {Buchalla}, \citenamefont {Cata}, \citenamefont {Rahn},\ and\ \citenamefont
  {Schlaffer}}]{Buchalla:2013wpa}%
  \BibitemOpen
  \bibfield  {author} {\bibinfo {author} {\bibfnamefont {G.}~\bibnamefont
  {Buchalla}}, \bibinfo {author} {\bibfnamefont {O.}~\bibnamefont {Cata}},
  \bibinfo {author} {\bibfnamefont {R.}~\bibnamefont {Rahn}},\ and\ \bibinfo
  {author} {\bibfnamefont {M.}~\bibnamefont {Schlaffer}},\ }\bibfield  {title}
  {\bibinfo {title} {{Effective Field Theory Analysis of New Physics in e+e-
  -\ensuremath{>} W+W- at a Linear Collider}},\ }\href
  {https://doi.org/10.1140/epjc/s10052-013-2589-1} {\bibfield  {journal}
  {\bibinfo  {journal} {Eur. Phys. J. C}\ }\textbf {\bibinfo {volume} {73}},\
  \bibinfo {pages} {2589} (\bibinfo {year} {2013})},\ \Eprint
  {https://arxiv.org/abs/1302.6481} {arXiv:1302.6481 [hep-ph]} \BibitemShut
  {NoStop}%
\bibitem [{\citenamefont {Choudhury}\ \emph {et~al.}(1999)\citenamefont
  {Choudhury}, \citenamefont {Kalinowski},\ and\ \citenamefont
  {Kulesza}}]{Choudhury:1999fz}%
  \BibitemOpen
  \bibfield  {author} {\bibinfo {author} {\bibfnamefont {D.}~\bibnamefont
  {Choudhury}}, \bibinfo {author} {\bibfnamefont {J.}~\bibnamefont
  {Kalinowski}},\ and\ \bibinfo {author} {\bibfnamefont {A.}~\bibnamefont
  {Kulesza}},\ }\bibfield  {title} {\bibinfo {title} {{CP violating anomalous
  $W W \gamma$ couplings in $e^{+} e^{-}$ collisions}},\ }\href
  {https://doi.org/10.1016/S0370-2693(99)00527-4} {\bibfield  {journal}
  {\bibinfo  {journal} {Phys. Lett. B}\ }\textbf {\bibinfo {volume} {457}},\
  \bibinfo {pages} {193} (\bibinfo {year} {1999})},\ \Eprint
  {https://arxiv.org/abs/hep-ph/9904215} {arXiv:hep-ph/9904215} \BibitemShut
  {NoStop}%
\bibitem [{\citenamefont {Subba}\ and\ \citenamefont
  {Singh}(2023{\natexlab{a}})}]{Subba:2022czw}%
  \BibitemOpen
  \bibfield  {author} {\bibinfo {author} {\bibfnamefont {A.}~\bibnamefont
  {Subba}}\ and\ \bibinfo {author} {\bibfnamefont {R.~K.}\ \bibnamefont
  {Singh}},\ }\bibfield  {title} {\bibinfo {title} {{Role of polarizations and
  spin-spin correlations of W's in e-e+\textrightarrow{}W-W+ at s=250\,\,GeV to
  probe anomalous W-W+Z/\ensuremath{\gamma} couplings}},\ }\href
  {https://doi.org/10.1103/PhysRevD.107.073004} {\bibfield  {journal} {\bibinfo
   {journal} {Phys. Rev. D}\ }\textbf {\bibinfo {volume} {107}},\ \bibinfo
  {pages} {073004} (\bibinfo {year} {2023}{\natexlab{a}})},\ \Eprint
  {https://arxiv.org/abs/2212.12973} {arXiv:2212.12973 [hep-ph]} \BibitemShut
  {NoStop}%
\bibitem [{\citenamefont {Subba}\ and\ \citenamefont
  {Singh}(2023{\natexlab{b}})}]{Subba:2023rpm}%
  \BibitemOpen
  \bibfield  {author} {\bibinfo {author} {\bibfnamefont {A.}~\bibnamefont
  {Subba}}\ and\ \bibinfo {author} {\bibfnamefont {R.~K.}\ \bibnamefont
  {Singh}},\ }\bibfield  {title} {\bibinfo {title} {{Study of anomalous
  $W^-W^+\gamma/Z$ couplings using polarizations and spin correlations in
  $e^-e^+\to W^-W^+$ with polarized beams}},\ }\href@noop {} {\  (\bibinfo
  {year} {2023}{\natexlab{b}})},\ \Eprint {https://arxiv.org/abs/2305.15106}
  {arXiv:2305.15106 [hep-ph]} \BibitemShut {NoStop}%
\bibitem [{\citenamefont {Rahaman}\ and\ \citenamefont
  {Singh}(2020{\natexlab{a}})}]{Rahaman:2019mnz}%
  \BibitemOpen
  \bibfield  {author} {\bibinfo {author} {\bibfnamefont {R.}~\bibnamefont
  {Rahaman}}\ and\ \bibinfo {author} {\bibfnamefont {R.~K.}\ \bibnamefont
  {Singh}},\ }\bibfield  {title} {\bibinfo {title} {{Probing the anomalous
  triple gauge boson couplings in $e^+e^-\to W^+W^-$ using $W$ polarizations
  with polarized beams}},\ }\href {https://doi.org/10.1103/PhysRevD.101.075044}
  {\bibfield  {journal} {\bibinfo  {journal} {Phys. Rev. D}\ }\textbf {\bibinfo
  {volume} {101}},\ \bibinfo {pages} {075044} (\bibinfo {year}
  {2020}{\natexlab{a}})},\ \Eprint {https://arxiv.org/abs/1909.05496}
  {arXiv:1909.05496 [hep-ph]} \BibitemShut {NoStop}%
\bibitem [{\citenamefont {Bian}\ \emph {et~al.}(2015)\citenamefont {Bian},
  \citenamefont {Shu},\ and\ \citenamefont {Zhang}}]{Bian:2015zha}%
  \BibitemOpen
  \bibfield  {author} {\bibinfo {author} {\bibfnamefont {L.}~\bibnamefont
  {Bian}}, \bibinfo {author} {\bibfnamefont {J.}~\bibnamefont {Shu}},\ and\
  \bibinfo {author} {\bibfnamefont {Y.}~\bibnamefont {Zhang}},\ }\bibfield
  {title} {\bibinfo {title} {{Prospects for Triple Gauge Coupling Measurements
  at Future Lepton Colliders and the 14 TeV LHC}},\ }\href
  {https://doi.org/10.1007/JHEP09(2015)206} {\bibfield  {journal} {\bibinfo
  {journal} {JHEP}\ }\textbf {\bibinfo {volume} {09}},\ \bibinfo {pages}
  {206}},\ \Eprint {https://arxiv.org/abs/1507.02238} {arXiv:1507.02238
  [hep-ph]} \BibitemShut {NoStop}%
\bibitem [{\citenamefont {Bian}\ \emph {et~al.}(2016)\citenamefont {Bian},
  \citenamefont {Shu},\ and\ \citenamefont {Zhang}}]{Bian:2016umx}%
  \BibitemOpen
  \bibfield  {author} {\bibinfo {author} {\bibfnamefont {L.}~\bibnamefont
  {Bian}}, \bibinfo {author} {\bibfnamefont {J.}~\bibnamefont {Shu}},\ and\
  \bibinfo {author} {\bibfnamefont {Y.}~\bibnamefont {Zhang}},\ }\bibfield
  {title} {\bibinfo {title} {{Triple gauge couplings at future hadron and
  lepton colliders}},\ }\href {https://doi.org/10.1142/S0217751X16440085}
  {\bibfield  {journal} {\bibinfo  {journal} {Int. J. Mod. Phys. A}\ }\textbf
  {\bibinfo {volume} {31}},\ \bibinfo {pages} {1644008} (\bibinfo {year}
  {2016})},\ \Eprint {https://arxiv.org/abs/1612.03888} {arXiv:1612.03888
  [hep-ph]} \BibitemShut {NoStop}%
\bibitem [{\citenamefont {Tizchang}\ and\ \citenamefont
  {Etesami}(2020)}]{Tizchang:2020tqs}%
  \BibitemOpen
  \bibfield  {author} {\bibinfo {author} {\bibfnamefont {S.}~\bibnamefont
  {Tizchang}}\ and\ \bibinfo {author} {\bibfnamefont {S.~M.}\ \bibnamefont
  {Etesami}},\ }\bibfield  {title} {\bibinfo {title} {{Pinning down the gauge
  boson couplings in WW\ensuremath{\gamma} production using forward proton
  tagging}},\ }\href {https://doi.org/10.1007/JHEP07(2020)191} {\bibfield
  {journal} {\bibinfo  {journal} {JHEP}\ }\textbf {\bibinfo {volume} {07}},\
  \bibinfo {pages} {191}},\ \Eprint {https://arxiv.org/abs/2004.12203}
  {arXiv:2004.12203 [hep-ph]} \BibitemShut {NoStop}%
\bibitem [{\citenamefont {Rahaman}\ and\ \citenamefont
  {Singh}(2020{\natexlab{b}})}]{Rahaman:2019lab}%
  \BibitemOpen
  \bibfield  {author} {\bibinfo {author} {\bibfnamefont {R.}~\bibnamefont
  {Rahaman}}\ and\ \bibinfo {author} {\bibfnamefont {R.~K.}\ \bibnamefont
  {Singh}},\ }\bibfield  {title} {\bibinfo {title} {{Unravelling the anomalous
  gauge boson couplings in $ZW^\pm$ production at the LHC and the role of
  spin-$1$ polarizations}},\ }\href {https://doi.org/10.1007/JHEP04(2020)075}
  {\bibfield  {journal} {\bibinfo  {journal} {JHEP}\ }\textbf {\bibinfo
  {volume} {04}},\ \bibinfo {pages} {075}},\ \Eprint
  {https://arxiv.org/abs/1911.03111} {arXiv:1911.03111 [hep-ph]} \BibitemShut
  {NoStop}%
\bibitem [{\citenamefont {Sirunyan}\ \emph {et~al.}(2021)\citenamefont
  {Sirunyan} \emph {et~al.}}]{CMS:2021foa}%
  \BibitemOpen
  \bibfield  {author} {\bibinfo {author} {\bibfnamefont {A.~M.}\ \bibnamefont
  {Sirunyan}} \emph {et~al.} (\bibinfo {collaboration} {CMS}),\ }\bibfield
  {title} {\bibinfo {title} {{Measurement of the W$\gamma$ Production Cross
  Section in Proton-Proton Collisions at $\sqrt {s}$=13\,\,TeV and Constraints
  on Effective Field Theory Coefficients}},\ }\href
  {https://doi.org/10.1103/PhysRevLett.126.252002} {\bibfield  {journal}
  {\bibinfo  {journal} {Phys. Rev. Lett.}\ }\textbf {\bibinfo {volume} {126}},\
  \bibinfo {pages} {252002} (\bibinfo {year} {2021})},\ \Eprint
  {https://arxiv.org/abs/2102.02283} {arXiv:2102.02283 [hep-ex]} \BibitemShut
  {NoStop}%
\bibitem [{\citenamefont {Ding}\ \emph {et~al.}(2020)\citenamefont {Ding},
  \citenamefont {Jones}, \citenamefont {Fuller}, \citenamefont {Martin},\ and\
  \citenamefont {Murray}}]{Ding:2019bkd}%
  \BibitemOpen
  \bibfield  {author} {\bibinfo {author} {\bibfnamefont {C.}~\bibnamefont
  {Ding}}, \bibinfo {author} {\bibfnamefont {E.}~\bibnamefont {Jones}},
  \bibinfo {author} {\bibfnamefont {B.}~\bibnamefont {Fuller}}, \bibinfo
  {author} {\bibfnamefont {A.}~\bibnamefont {Martin}},\ and\ \bibinfo {author}
  {\bibfnamefont {W.}~\bibnamefont {Murray}},\ }\bibfield  {title} {\bibinfo
  {title} {{Electroweak-QCD interference in hadronic vector bosons decays at
  the LHC}},\ }\href {https://doi.org/10.1140/epjc/s10052-020-7729-9}
  {\bibfield  {journal} {\bibinfo  {journal} {Eur. Phys. J. C}\ }\textbf
  {\bibinfo {volume} {80}},\ \bibinfo {pages} {176} (\bibinfo {year} {2020})},\
  \Eprint {https://arxiv.org/abs/1908.08330} {arXiv:1908.08330 [hep-ph]}
  \BibitemShut {NoStop}%
\bibitem [{\citenamefont {Baur}\ \emph {et~al.}(1989)\citenamefont {Baur},
  \citenamefont {Glover},\ and\ \citenamefont {Martin}}]{Baur:1989qt}%
  \BibitemOpen
  \bibfield  {author} {\bibinfo {author} {\bibfnamefont {U.}~\bibnamefont
  {Baur}}, \bibinfo {author} {\bibfnamefont {E.~W.~N.}\ \bibnamefont
  {Glover}},\ and\ \bibinfo {author} {\bibfnamefont {A.~D.}\ \bibnamefont
  {Martin}},\ }\bibfield  {title} {\bibinfo {title} {{Electroweak Interference
  Effects in Two Jet Production at $p \bar{p}$ Colliders}},\ }\href
  {https://doi.org/10.1016/0370-2693(89)90452-8} {\bibfield  {journal}
  {\bibinfo  {journal} {Phys. Lett. B}\ }\textbf {\bibinfo {volume} {232}},\
  \bibinfo {pages} {519} (\bibinfo {year} {1989})}\BibitemShut {NoStop}%
\bibitem [{\citenamefont {Ranft}\ and\ \citenamefont
  {Ranft}(1979)}]{RANFT1979122}%
  \BibitemOpen
  \bibfield  {author} {\bibinfo {author} {\bibfnamefont {G.}~\bibnamefont
  {Ranft}}\ and\ \bibinfo {author} {\bibfnamefont {J.}~\bibnamefont {Ranft}},\
  }\bibfield  {title} {\bibinfo {title} {Qcd-weak interference and predictions
  for vector boson signals in hadronic jet cross sections in polarized and
  unpolarized pp and pp collisions},\ }\href
  {https://doi.org/https://doi.org/10.1016/0370-2693(79)90034-0} {\bibfield
  {journal} {\bibinfo  {journal} {Physics Letters B}\ }\textbf {\bibinfo
  {volume} {87}},\ \bibinfo {pages} {122} (\bibinfo {year} {1979})}\BibitemShut
  {NoStop}%
\bibitem [{\citenamefont {Alwall}\ \emph {et~al.}(2014)\citenamefont {Alwall},
  \citenamefont {Frederix}, \citenamefont {Frixione}, \citenamefont {Hirschi},
  \citenamefont {Maltoni}, \citenamefont {Mattelaer}, \citenamefont {Shao},
  \citenamefont {Stelzer}, \citenamefont {Torrielli},\ and\ \citenamefont
  {Zaro}}]{Alwall:2014hca}%
  \BibitemOpen
  \bibfield  {author} {\bibinfo {author} {\bibfnamefont {J.}~\bibnamefont
  {Alwall}}, \bibinfo {author} {\bibfnamefont {R.}~\bibnamefont {Frederix}},
  \bibinfo {author} {\bibfnamefont {S.}~\bibnamefont {Frixione}}, \bibinfo
  {author} {\bibfnamefont {V.}~\bibnamefont {Hirschi}}, \bibinfo {author}
  {\bibfnamefont {F.}~\bibnamefont {Maltoni}}, \bibinfo {author} {\bibfnamefont
  {O.}~\bibnamefont {Mattelaer}}, \bibinfo {author} {\bibfnamefont {H.~S.}\
  \bibnamefont {Shao}}, \bibinfo {author} {\bibfnamefont {T.}~\bibnamefont
  {Stelzer}}, \bibinfo {author} {\bibfnamefont {P.}~\bibnamefont {Torrielli}},\
  and\ \bibinfo {author} {\bibfnamefont {M.}~\bibnamefont {Zaro}},\ }\bibfield
  {title} {\bibinfo {title} {{The automated computation of tree-level and
  next-to-leading order differential cross sections, and their matching to
  parton shower simulations}},\ }\href
  {https://doi.org/10.1007/JHEP07(2014)079} {\bibfield  {journal} {\bibinfo
  {journal} {JHEP}\ }\textbf {\bibinfo {volume} {07}},\ \bibinfo {pages}
  {079}},\ \Eprint {https://arxiv.org/abs/1405.0301} {arXiv:1405.0301 [hep-ph]}
  \BibitemShut {NoStop}%
\bibitem [{Beh(2013)}]{Behnke:2013xla}%
  \BibitemOpen
  \bibfield  {title} {\bibinfo {title} {{The International Linear Collider
  Technical Design Report - Volume 1: Executive Summary}},\ }\href@noop {} {\
  (\bibinfo {year} {2013})},\ \Eprint {https://arxiv.org/abs/1306.6327}
  {arXiv:1306.6327 [physics.acc-ph]} \BibitemShut {NoStop}%
\bibitem [{Aic(2012)}]{Aicheler:2012bya}%
  \BibitemOpen
  \bibfield  {title} {\bibinfo {title} {{A Multi-TeV Linear Collider Based on
  CLIC Technology}: {CLIC Conceptual Design Report}}\ }\href
  {https://doi.org/10.5170/CERN-2012-007} {10.5170/CERN-2012-007} (\bibinfo
  {year} {2012})\BibitemShut {NoStop}%
\bibitem [{Lin(2012)}]{Linssen:2012hp}%
  \BibitemOpen
  \bibfield  {title} {\bibinfo {title} {{Physics and Detectors at CLIC: CLIC
  Conceptual Design Report}}\ }\href {https://doi.org/10.5170/CERN-2012-003}
  {10.5170/CERN-2012-003} (\bibinfo {year} {2012}),\ \Eprint
  {https://arxiv.org/abs/1202.5940} {arXiv:1202.5940 [physics.ins-det]}
  \BibitemShut {NoStop}%
\bibitem [{\citenamefont {Lebrun}\ \emph {et~al.}(2012)\citenamefont {Lebrun},
  \citenamefont {Linssen}, \citenamefont {Lucaci-Timoce}, \citenamefont
  {Schulte}, \citenamefont {Simon}, \citenamefont {Stapnes}, \citenamefont
  {Toge}, \citenamefont {Weerts},\ and\ \citenamefont {Wells}}]{Lebrun:2012hj}%
  \BibitemOpen
  \bibfield  {author} {\bibinfo {author} {\bibfnamefont {P.}~\bibnamefont
  {Lebrun}}, \bibinfo {author} {\bibfnamefont {L.}~\bibnamefont {Linssen}},
  \bibinfo {author} {\bibfnamefont {A.}~\bibnamefont {Lucaci-Timoce}}, \bibinfo
  {author} {\bibfnamefont {D.}~\bibnamefont {Schulte}}, \bibinfo {author}
  {\bibfnamefont {F.}~\bibnamefont {Simon}}, \bibinfo {author} {\bibfnamefont
  {S.}~\bibnamefont {Stapnes}}, \bibinfo {author} {\bibfnamefont
  {N.}~\bibnamefont {Toge}}, \bibinfo {author} {\bibfnamefont {H.}~\bibnamefont
  {Weerts}},\ and\ \bibinfo {author} {\bibfnamefont {J.}~\bibnamefont
  {Wells}},\ }\bibfield  {title} {\bibinfo {title} {{The CLIC Programme:
  Towards a Staged e+e- Linear Collider Exploring the Terascale : CLIC
  Conceptual Design Report}}\ }\href {https://doi.org/10.5170/CERN-2012-005}
  {10.5170/CERN-2012-005} (\bibinfo {year} {2012}),\ \Eprint
  {https://arxiv.org/abs/1209.2543} {arXiv:1209.2543 [physics.ins-det]}
  \BibitemShut {NoStop}%
\bibitem [{\citenamefont {Abada}\ \emph {et~al.}(2019)\citenamefont {Abada}
  \emph {et~al.}}]{FCC:2018evy}%
  \BibitemOpen
  \bibfield  {author} {\bibinfo {author} {\bibfnamefont {A.}~\bibnamefont
  {Abada}} \emph {et~al.} (\bibinfo {collaboration} {FCC}),\ }\bibfield
  {title} {\bibinfo {title} {{FCC-ee: The Lepton Collider}: {Future Circular
  Collider Conceptual Design Report Volume 2}},\ }\href
  {https://doi.org/10.1140/epjst/e2019-900045-4} {\bibfield  {journal}
  {\bibinfo  {journal} {Eur. Phys. J. ST}\ }\textbf {\bibinfo {volume} {228}},\
  \bibinfo {pages} {261} (\bibinfo {year} {2019})}\BibitemShut {NoStop}%
\bibitem [{CEP(2018)}]{CEPCStudyGroup:2018rmc}%
  \BibitemOpen
  \bibfield  {title} {\bibinfo {title} {{CEPC Conceptual Design Report: Volume
  1 - Accelerator}},\ }\href@noop {} {\  (\bibinfo {year} {2018})},\ \Eprint
  {https://arxiv.org/abs/1809.00285} {arXiv:1809.00285 [physics.acc-ph]}
  \BibitemShut {NoStop}%
\bibitem [{\citenamefont {Li}(2023)}]{Li:2023kjt}%
  \BibitemOpen
  \bibfield  {author} {\bibinfo {author} {\bibfnamefont {Y.}~\bibnamefont
  {Li}},\ }\bibfield  {title} {\bibinfo {title} {{CEPC Accelerator TDR Status
  Overview}},\ }\href {https://doi.org/10.18429/JACoW-eeFACT2022-MOXAT0105}
  {\bibfield  {journal} {\bibinfo  {journal} {JACoW}\ }\textbf {\bibinfo
  {volume} {eeFACT2022}},\ \bibinfo {pages} {14} (\bibinfo {year}
  {2023})}\BibitemShut {NoStop}%
\bibitem [{\citenamefont {Boudjema}\ and\ \citenamefont
  {Singh}(2009)}]{Boudjema:2009fz}%
  \BibitemOpen
  \bibfield  {author} {\bibinfo {author} {\bibfnamefont {F.}~\bibnamefont
  {Boudjema}}\ and\ \bibinfo {author} {\bibfnamefont {R.~K.}\ \bibnamefont
  {Singh}},\ }\bibfield  {title} {\bibinfo {title} {{A Model independent spin
  analysis of fundamental particles using azimuthal asymmetries}},\ }\href
  {https://doi.org/10.1088/1126-6708/2009/07/028} {\bibfield  {journal}
  {\bibinfo  {journal} {JHEP}\ }\textbf {\bibinfo {volume} {07}},\ \bibinfo
  {pages} {028}},\ \Eprint {https://arxiv.org/abs/0903.4705} {arXiv:0903.4705
  [hep-ph]} \BibitemShut {NoStop}%
\bibitem [{\citenamefont {Rahaman}\ and\ \citenamefont
  {Singh}(2022)}]{Rahaman:2021fcz}%
  \BibitemOpen
  \bibfield  {author} {\bibinfo {author} {\bibfnamefont {R.}~\bibnamefont
  {Rahaman}}\ and\ \bibinfo {author} {\bibfnamefont {R.~K.}\ \bibnamefont
  {Singh}},\ }\bibfield  {title} {\bibinfo {title} {{Breaking down the entire
  spectrum of spin correlations of a pair of particles involving fermions and
  gauge bosons}},\ }\href {https://doi.org/10.1016/j.nuclphysb.2022.115984}
  {\bibfield  {journal} {\bibinfo  {journal} {Nucl. Phys. B}\ }\textbf
  {\bibinfo {volume} {984}},\ \bibinfo {pages} {115984} (\bibinfo {year}
  {2022})},\ \Eprint {https://arxiv.org/abs/2109.09345} {arXiv:2109.09345
  [hep-ph]} \BibitemShut {NoStop}%
\bibitem [{\citenamefont {Field}\ and\ \citenamefont
  {Feynman}(1978)}]{Field:1977fa}%
  \BibitemOpen
  \bibfield  {author} {\bibinfo {author} {\bibfnamefont {R.~D.}\ \bibnamefont
  {Field}}\ and\ \bibinfo {author} {\bibfnamefont {R.~P.}\ \bibnamefont
  {Feynman}},\ }\bibfield  {title} {\bibinfo {title} {{A Parametrization of the
  Properties of Quark Jets}},\ }\href
  {https://doi.org/10.1016/0550-3213(78)90015-9} {\bibfield  {journal}
  {\bibinfo  {journal} {Nucl. Phys. B}\ }\textbf {\bibinfo {volume} {136}},\
  \bibinfo {pages} {1} (\bibinfo {year} {1978})}\BibitemShut {NoStop}%
\bibitem [{\citenamefont {Waalewijn}(2012)}]{Waalewijn:2012sv}%
  \BibitemOpen
  \bibfield  {author} {\bibinfo {author} {\bibfnamefont {W.~J.}\ \bibnamefont
  {Waalewijn}},\ }\bibfield  {title} {\bibinfo {title} {{Calculating the Charge
  of a Jet}},\ }\href {https://doi.org/10.1103/PhysRevD.86.094030} {\bibfield
  {journal} {\bibinfo  {journal} {Phys. Rev. D}\ }\textbf {\bibinfo {volume}
  {86}},\ \bibinfo {pages} {094030} (\bibinfo {year} {2012})},\ \Eprint
  {https://arxiv.org/abs/1209.3019} {arXiv:1209.3019 [hep-ph]} \BibitemShut
  {NoStop}%
\bibitem [{\citenamefont {Chang}\ \emph {et~al.}(2013)\citenamefont {Chang},
  \citenamefont {Procura}, \citenamefont {Thaler},\ and\ \citenamefont
  {Waalewijn}}]{Chang:2013rca}%
  \BibitemOpen
  \bibfield  {author} {\bibinfo {author} {\bibfnamefont {H.-M.}\ \bibnamefont
  {Chang}}, \bibinfo {author} {\bibfnamefont {M.}~\bibnamefont {Procura}},
  \bibinfo {author} {\bibfnamefont {J.}~\bibnamefont {Thaler}},\ and\ \bibinfo
  {author} {\bibfnamefont {W.~J.}\ \bibnamefont {Waalewijn}},\ }\bibfield
  {title} {\bibinfo {title} {{Calculating Track-Based Observables for the
  LHC}},\ }\href {https://doi.org/10.1103/PhysRevLett.111.102002} {\bibfield
  {journal} {\bibinfo  {journal} {Phys. Rev. Lett.}\ }\textbf {\bibinfo
  {volume} {111}},\ \bibinfo {pages} {102002} (\bibinfo {year} {2013})},\
  \Eprint {https://arxiv.org/abs/1303.6637} {arXiv:1303.6637 [hep-ph]}
  \BibitemShut {NoStop}%
\bibitem [{\citenamefont {Kang}\ \emph {et~al.}(2023)\citenamefont {Kang},
  \citenamefont {Larkoski},\ and\ \citenamefont {Yang}}]{Kang:2023ptt}%
  \BibitemOpen
  \bibfield  {author} {\bibinfo {author} {\bibfnamefont {Z.-B.}\ \bibnamefont
  {Kang}}, \bibinfo {author} {\bibfnamefont {A.~J.}\ \bibnamefont {Larkoski}},\
  and\ \bibinfo {author} {\bibfnamefont {J.}~\bibnamefont {Yang}},\ }\bibfield
  {title} {\bibinfo {title} {{Towards a Nonperturbative Formulation of the Jet
  Charge}},\ }\href {https://doi.org/10.1103/PhysRevLett.130.151901} {\bibfield
   {journal} {\bibinfo  {journal} {Phys. Rev. Lett.}\ }\textbf {\bibinfo
  {volume} {130}},\ \bibinfo {pages} {151901} (\bibinfo {year} {2023})},\
  \Eprint {https://arxiv.org/abs/2301.09649} {arXiv:2301.09649 [hep-ph]}
  \BibitemShut {NoStop}%
\bibitem [{\citenamefont {Aaltonen}\ \emph {et~al.}(2022)\citenamefont
  {Aaltonen} \emph {et~al.}}]{CDF:2022hxs}%
  \BibitemOpen
  \bibfield  {author} {\bibinfo {author} {\bibfnamefont {T.}~\bibnamefont
  {Aaltonen}} \emph {et~al.} (\bibinfo {collaboration} {CDF}),\ }\bibfield
  {title} {\bibinfo {title} {{High-precision measurement of the $W$ boson mass
  with the CDF II detector}},\ }\href {https://doi.org/10.1126/science.abk1781}
  {\bibfield  {journal} {\bibinfo  {journal} {Science}\ }\textbf {\bibinfo
  {volume} {376}},\ \bibinfo {pages} {170} (\bibinfo {year}
  {2022})}\BibitemShut {NoStop}%
\bibitem [{\citenamefont {Albahri}\ \emph {et~al.}(2021)\citenamefont {Albahri}
  \emph {et~al.}}]{Muong-2:2021vma}%
  \BibitemOpen
  \bibfield  {author} {\bibinfo {author} {\bibfnamefont {T.}~\bibnamefont
  {Albahri}} \emph {et~al.} (\bibinfo {collaboration} {Muon g-2}),\ }\bibfield
  {title} {\bibinfo {title} {{Measurement of the anomalous precession frequency
  of the muon in the Fermilab Muon $g−2$ Experiment}},\ }\href
  {https://doi.org/10.1103/PhysRevD.103.072002} {\bibfield  {journal} {\bibinfo
   {journal} {Phys. Rev. D}\ }\textbf {\bibinfo {volume} {103}},\ \bibinfo
  {pages} {072002} (\bibinfo {year} {2021})},\ \Eprint
  {https://arxiv.org/abs/2104.03247} {arXiv:2104.03247 [hep-ex]} \BibitemShut
  {NoStop}%
\bibitem [{\citenamefont {Abi}\ \emph {et~al.}(2021)\citenamefont {Abi} \emph
  {et~al.}}]{Muong-2:2021ojo}%
  \BibitemOpen
  \bibfield  {author} {\bibinfo {author} {\bibfnamefont {B.}~\bibnamefont
  {Abi}} \emph {et~al.} (\bibinfo {collaboration} {Muon g-2}),\ }\bibfield
  {title} {\bibinfo {title} {{Measurement of the Positive Muon Anomalous
  Magnetic Moment to 0.46 ppm}},\ }\href
  {https://doi.org/10.1103/PhysRevLett.126.141801} {\bibfield  {journal}
  {\bibinfo  {journal} {Phys. Rev. Lett.}\ }\textbf {\bibinfo {volume} {126}},\
  \bibinfo {pages} {141801} (\bibinfo {year} {2021})},\ \Eprint
  {https://arxiv.org/abs/2104.03281} {arXiv:2104.03281 [hep-ex]} \BibitemShut
  {NoStop}%
\bibitem [{\citenamefont {Appelquist}\ and\ \citenamefont
  {Carazzone}(1975)}]{Appelquist:1974tg}%
  \BibitemOpen
  \bibfield  {author} {\bibinfo {author} {\bibfnamefont {T.}~\bibnamefont
  {Appelquist}}\ and\ \bibinfo {author} {\bibfnamefont {J.}~\bibnamefont
  {Carazzone}},\ }\bibfield  {title} {\bibinfo {title} {{Infrared Singularities
  and Massive Fields}},\ }\href {https://doi.org/10.1103/PhysRevD.11.2856}
  {\bibfield  {journal} {\bibinfo  {journal} {Phys. Rev. D}\ }\textbf {\bibinfo
  {volume} {11}},\ \bibinfo {pages} {2856} (\bibinfo {year}
  {1975})}\BibitemShut {NoStop}%
\bibitem [{\citenamefont {Sirunyan}\ \emph
  {et~al.}(2019{\natexlab{b}})\citenamefont {Sirunyan} \emph
  {et~al.}}]{CMS:2019ppl}%
  \BibitemOpen
  \bibfield  {author} {\bibinfo {author} {\bibfnamefont {A.~M.}\ \bibnamefont
  {Sirunyan}} \emph {et~al.} (\bibinfo {collaboration} {CMS}),\ }\bibfield
  {title} {\bibinfo {title} {{Search for anomalous triple gauge couplings in WW
  and WZ production in lepton + jet events in proton-proton collisions at
  $\sqrt{s} =$ 13 TeV}},\ }\href {https://doi.org/10.1007/JHEP12(2019)062}
  {\bibfield  {journal} {\bibinfo  {journal} {JHEP}\ }\textbf {\bibinfo
  {volume} {12}},\ \bibinfo {pages} {062}},\ \Eprint
  {https://arxiv.org/abs/1907.08354} {arXiv:1907.08354 [hep-ex]} \BibitemShut
  {NoStop}%
\bibitem [{\citenamefont {Aghanim}\ \emph {et~al.}(2020)\citenamefont {Aghanim}
  \emph {et~al.}}]{Planck:2018vyg}%
  \BibitemOpen
  \bibfield  {author} {\bibinfo {author} {\bibfnamefont {N.}~\bibnamefont
  {Aghanim}} \emph {et~al.} (\bibinfo {collaboration} {Planck}),\ }\bibfield
  {title} {\bibinfo {title} {{Planck 2018 results. VI. Cosmological
  parameters}},\ }\href {https://doi.org/10.1051/0004-6361/201833910}
  {\bibfield  {journal} {\bibinfo  {journal} {Astron. Astrophys.}\ }\textbf
  {\bibinfo {volume} {641}},\ \bibinfo {pages} {A6} (\bibinfo {year} {2020})},\
  \bibinfo {note} {[Erratum: Astron.Astrophys. 652, C4 (2021)]},\ \Eprint
  {https://arxiv.org/abs/1807.06209} {arXiv:1807.06209 [astro-ph.CO]}
  \BibitemShut {NoStop}%
\bibitem [{\citenamefont {Aguillard}\ \emph {et~al.}(2023)\citenamefont
  {Aguillard} \emph {et~al.}}]{Muong-2:2023cdq}%
  \BibitemOpen
  \bibfield  {author} {\bibinfo {author} {\bibfnamefont {D.~P.}\ \bibnamefont
  {Aguillard}} \emph {et~al.} (\bibinfo {collaboration} {Muon g-2}),\
  }\bibfield  {title} {\bibinfo {title} {{Measurement of the Positive Muon
  Anomalous Magnetic Moment to 0.20 ppm}},\ }\href@noop {} {\  (\bibinfo {year}
  {2023})},\ \Eprint {https://arxiv.org/abs/2308.06230} {arXiv:2308.06230
  [hep-ex]} \BibitemShut {NoStop}%
\bibitem [{\citenamefont {Liu}\ and\ \citenamefont {Segre}(1994)}]{Liu:1993am}%
  \BibitemOpen
  \bibfield  {author} {\bibinfo {author} {\bibfnamefont {J.}~\bibnamefont
  {Liu}}\ and\ \bibinfo {author} {\bibfnamefont {G.}~\bibnamefont {Segre}},\
  }\bibfield  {title} {\bibinfo {title} {{Baryon asymmetry of the universe and
  large lepton asymmetries}},\ }\href
  {https://doi.org/10.1016/0370-2693(94)91375-7} {\bibfield  {journal}
  {\bibinfo  {journal} {Phys. Lett. B}\ }\textbf {\bibinfo {volume} {338}},\
  \bibinfo {pages} {259} (\bibinfo {year} {1994})}\BibitemShut {NoStop}%
\bibitem [{\citenamefont {Frixione}\ \emph {et~al.}(2008)\citenamefont
  {Frixione}, \citenamefont {Laenen}, \citenamefont {Motylinski}, \citenamefont
  {Webber},\ and\ \citenamefont {White}}]{Frixione:2008yi}%
  \BibitemOpen
  \bibfield  {author} {\bibinfo {author} {\bibfnamefont {S.}~\bibnamefont
  {Frixione}}, \bibinfo {author} {\bibfnamefont {E.}~\bibnamefont {Laenen}},
  \bibinfo {author} {\bibfnamefont {P.}~\bibnamefont {Motylinski}}, \bibinfo
  {author} {\bibfnamefont {B.~R.}\ \bibnamefont {Webber}},\ and\ \bibinfo
  {author} {\bibfnamefont {C.~D.}\ \bibnamefont {White}},\ }\bibfield  {title}
  {\bibinfo {title} {{Single-top hadroproduction in association with a W
  boson}},\ }\href {https://doi.org/10.1088/1126-6708/2008/07/029} {\bibfield
  {journal} {\bibinfo  {journal} {JHEP}\ }\textbf {\bibinfo {volume} {07}},\
  \bibinfo {pages} {029}},\ \Eprint {https://arxiv.org/abs/0805.3067}
  {arXiv:0805.3067 [hep-ph]} \BibitemShut {NoStop}%
\bibitem [{\citenamefont {Wudka}(1994)}]{Wudka:1994ny}%
  \BibitemOpen
  \bibfield  {author} {\bibinfo {author} {\bibfnamefont {J.}~\bibnamefont
  {Wudka}},\ }\bibfield  {title} {\bibinfo {title} {{Electroweak effective
  Lagrangians}},\ }\href {https://doi.org/10.1142/S0217751X94000959} {\bibfield
   {journal} {\bibinfo  {journal} {Int. J. Mod. Phys. A}\ }\textbf {\bibinfo
  {volume} {9}},\ \bibinfo {pages} {2301} (\bibinfo {year} {1994})},\ \Eprint
  {https://arxiv.org/abs/hep-ph/9406205} {arXiv:hep-ph/9406205} \BibitemShut
  {NoStop}%
\bibitem [{\citenamefont {Aguilar-Saavedra}\ \emph {et~al.}(2017)\citenamefont
  {Aguilar-Saavedra}, \citenamefont {Bernab\'eu}, \citenamefont {Mitsou},\ and\
  \citenamefont {Segarra}}]{Aguilar-Saavedra:2017zkn}%
  \BibitemOpen
  \bibfield  {author} {\bibinfo {author} {\bibfnamefont {J.~A.}\ \bibnamefont
  {Aguilar-Saavedra}}, \bibinfo {author} {\bibfnamefont {J.}~\bibnamefont
  {Bernab\'eu}}, \bibinfo {author} {\bibfnamefont {V.~A.}\ \bibnamefont
  {Mitsou}},\ and\ \bibinfo {author} {\bibfnamefont {A.}~\bibnamefont
  {Segarra}},\ }\bibfield  {title} {\bibinfo {title} {{The Z boson spin
  observables as messengers of new physics}},\ }\href
  {https://doi.org/10.1140/epjc/s10052-017-4795-8} {\bibfield  {journal}
  {\bibinfo  {journal} {Eur. Phys. J. C}\ }\textbf {\bibinfo {volume} {77}},\
  \bibinfo {pages} {234} (\bibinfo {year} {2017})},\ \Eprint
  {https://arxiv.org/abs/1701.03115} {arXiv:1701.03115 [hep-ph]} \BibitemShut
  {NoStop}%
\bibitem [{\citenamefont {Aguilar-Saavedra}\ and\ \citenamefont
  {Bernabeu}(2016)}]{Aguilar-Saavedra:2015yza}%
  \BibitemOpen
  \bibfield  {author} {\bibinfo {author} {\bibfnamefont {J.~A.}\ \bibnamefont
  {Aguilar-Saavedra}}\ and\ \bibinfo {author} {\bibfnamefont {J.}~\bibnamefont
  {Bernabeu}},\ }\bibfield  {title} {\bibinfo {title} {{Breaking down the
  entire W boson spin observables from its decay}},\ }\href
  {https://doi.org/10.1103/PhysRevD.93.011301} {\bibfield  {journal} {\bibinfo
  {journal} {Phys. Rev. D}\ }\textbf {\bibinfo {volume} {93}},\ \bibinfo
  {pages} {011301} (\bibinfo {year} {2016})},\ \Eprint
  {https://arxiv.org/abs/1508.04592} {arXiv:1508.04592 [hep-ph]} \BibitemShut
  {NoStop}%
\end{thebibliography}%
	
\end{document}